\documentclass[aps,epsfig,floats,pre,twocolumn,showpacs]{revtex4-1}
\usepackage{graphicx}
\usepackage[latin2]{inputenc}
\usepackage{amsmath}
\usepackage{amssymb}
\usepackage{epsfig}
\usepackage{bm}
\usepackage{color}

\begin{document}

\title{Persistent random walk approach to anomalous transport of self-propelled particles}

\author{Zeinab Sadjadi}
\email{sadjadi@lusi.uni-sb.de}
\affiliation{Department of Theoretical Physics, Saarland 
University, D-66041 Saarbr\"ucken, Germany}

\author{M.\ Reza Shaebani}
\email{shaebani@lusi.uni-sb.de}
\affiliation{Department of Theoretical Physics, Saarland 
University, D-66041 Saarbr\"ucken, Germany}

\author{Heiko Rieger}
\affiliation{Department of Theoretical Physics, Saarland 
University, D-66041 Saarbr\"ucken, Germany}

\author{Ludger Santen}
\affiliation{Department of Theoretical Physics, Saarland 
University, D-66041 Saarbr\"ucken, Germany}
\date{\today}

\begin{abstract}
The motion of self-propelled particles is modeled as a persistent 
random walk. An analytical framework is developed that allows the 
derivation of exact expressions for the time evolution of arbitrary 
moments of the persistent walk's displacement. It is shown 
that the interplay of step length and turning angle distributions 
and self-propulsion produces various signs of anomalous diffusion 
at short time scales and asymptotically a normal diffusion behavior 
with a broad range of diffusion coefficients. The crossover from 
the anomalous short time behavior to the asymptotic diffusion 
regime is studied and the parameter dependencies of the crossover 
time are discussed. Higher moments of the displacement distribution 
are calculated and analytical expressions for the time evolution 
of the skewness and the kurtosis of the distribution are presented.
\end{abstract}

\pacs{87.16.Uv, 05.40.-a, 02.50.-r, 87.16.Ka, 87.16.Nn}

\maketitle

\section{Introduction}
\label{Introduction}
Self-propelled particles undergo active Brownian motion by 
consuming energy, obtained either from internal or external 
sources. Examples range from the transport of motor proteins 
on cytoskeletal filaments \cite{Ross08} which is a biologically 
relevant system, to the motion of self-motile colloidal 
particles \cite{Howse07} as a nonliving realization. The 
particles are powered by the hydrolysis of ATP in the former 
case, whereas they use a chemical reaction catalyzed on their 
surface to swim in the latter example. The directed propulsion 
subject to fluctuations has been described by persistent 
random walks \cite{Howse07,Peruani07,Shaebani14}, where a 
tendency to move along the previous direction is taken into 
account. A strong self-propulsion overcomes the stochastic 
fluctuations and directs the motion, which renders it ballistic 
for short time scales \cite{Howse07,Shaebani14}. Nevertheless, 
the interplay between self-propulsion and random motion in 
general may lead to various scenarios of anomalous diffusive 
dynamics on varying time scales \cite{Shaebani14}. The influence 
of self-propulsion diminishes over time and eventually a 
crossover to an asymptotic diffusive regime occurs.

Even in the absence of self-propulsion, the stochastic 
motion of particle may remain complicated because in 
general a random walker can perform steps with arbitrary 
turning angles and variable step lengths. Moreover, there 
can be a relation between the step size and the turning angle 
of each step. Generalized random walks had been studied 
e.g.\ in the context of animal and cell movements as a 
Markovian process \cite{OldRefs,Codling08}, i.e.\ 
by considering the motion as a series of independent draws 
from the step-length and turning-angle distributions for 
each step. While the focus of prior studies has been more 
on the asymptotic diffusion coefficient of such random walks, 
the short-time behavior is neither thoroughly investigated 
nor completely understood. 

In the persistent random walk model that we study here a particle 
moves straight in continuous space in a randomly chosen direction 
over a randomly chosen distance and then changes direction by a 
randomly chosen turning angle. The typical trajectories of the 
random walker depend strongly on the chosen turning angle and 
distance distributions. For small values of the angular 
change the new direction will be strongly correlated with the old 
direction, introducing a directional memory into the model without 
changing the Markov property of the process. The emerging 
intermittent directional bias is controlled by the characteristics 
(mean, width, asymmetry, etc.) of the turning angle distribution 
and the probability with which the direction is unchanged, i.e.\ 
the processivity. The bias decays with time after a few turns 
and the directions of the particle motion become asymptotically 
randomized. 

Memory effects have also been included in other random walk 
models. In fractional Brownian motion \cite{FBM}, sub or 
superdiffusive motion is observed asymptotically. In continuous 
time random walk (CTRW) models \cite{CTRW}, for example, true 
non-Markovian effects can be implemented via broad waiting time 
distributions which lead to a subdiffusive dynamics of the 
walker. Although these approaches are conceptually very exciting, 
a direct comparison to experimental results is sometimes 
difficult. In case of intracellular transport, for example, 
spatial confinement as well as a finite observation time imply 
that the asymptotic behavior of the walker is not accessible. 

Here, we develop a general analytical framework to study 
persistent random walks over the whole range of time scales. 
The goal is to clarify and disentangle the combined effects 
of self-propulsion $p$ and the stepping strategy of the 
walker, consisting of its step-length $\mathcal{F}(\ell)$ and 
turning-angle $R(\phi)$ distributions. The method enables 
us to analytically determine the time evolution of arbitrary 
moments of displacement. Using this approach, the second 
moment, i.e.\ the mean square displacement, has been 
recently analyzed \cite{Shaebani14} revealing a variety 
of signatures of anomalous diffusion on short time scales 
even in the absence of viscoelasticity, traps, or overcrowding. 
These elements were frequently identified in the nature 
of the biological environments and received considerable 
attention as the possible sources of subdiffusion \cite{Reviews}.

An alternative way to interpret the model and its outcome 
is to consider the motion on complex networks, such as motor 
proteins on the cytoskeleton. These motors have an effective 
processivity $p$ (i.e.\ a tendency to move along the same 
filament \cite{ProcesivityRefs1,Vershinin07,ProcesivityRefs2}) 
and may switch to a new filament at the intersections of 
the network. From a coarse-grained perspective, one can 
describe such motion by a persistent random walk on the nodes 
of the network, thus, $\mathcal{F}(\ell)$ and $R(\phi)$ 
represent the distributions of the segment-length $\ell$ 
between neighboring intersections and the angle $\phi$ 
between intersecting filaments, respectively. Note that 
the cytoskeleton is a dynamic network due to the underlying 
growth and shrinkage of filaments, thus, the structure 
on which the transport takes place often changes. This 
justifies the relevance of the proposed stochastic approach, 
where the network structure is always implicitly given.  

Our analytical approach provides a recipe to obtain any 
arbitrary moment of displacement, yet we extend the 
calculations to the third and fourth moments, which 
are of particular interest, e.g.\ in the evaluation 
of the skewness and kurtosis, or to obtain the variance 
$\sigma_{r^2}{=}\langle r^4 \rangle {-} \langle r^2 
\rangle^2$ (and thus the standard error) of the mean 
square displacement (MSD) of a persistent random walker. 
$\sigma_{r^2}$ is also a useful quantity for estimating 
the first moment $\langle r \rangle$ of the net 
displacement: In the absence of an exact analytical 
solution for $\langle r \rangle$, an approximate 
expression $\langle r \rangle \approx \sqrt{\langle 
r^2 \rangle} (1{-}\frac18 \frac{\sigma_{r^2}}{\langle 
r^2 \rangle^2})$ was proposed \cite{McCulloch89} by means 
of a Taylor expansion of the square root function. The 
asymptotic behavior is however shown to follow $\langle 
r \rangle = \frac12 \sqrt{\pi \langle r^2 \rangle}$ 
\cite{Bovet88}.

We first introduce the general approach to obtain arbitrary 
moments of displacement. Next, an analytical expression is 
derived for the special case of the mean square displacement 
in two dimensions. This part is the full exposition and 
expansion of the results presented in \cite{Shaebani14}. 
Then we clarify the similarities and differences between 
the results in two and three dimensions for persistent 
walks which are symmetric around the previous direction 
of motion, and briefly discuss the case of asymmetric 
turning-angle distributions in two dimensions. Finally, 
the calculations are extended to higher moments and 
cumulants of displacement, and the probability density 
of the position of the random walker and its time evolution 
are investigated. The paper is organized in the following 
manner: Section~\ref{Model-2D} contains the description 
of the master equation formalism for the persistent 
motion of random walkers in two dimensions. In 
Sec.~\ref{Model-3D}, we discuss symmetric persistent 
motions in three dimensions. The analytical predictions 
for MSD are compared with simulation results in 
Sec.~\ref{Results}. The calculations are extended to 
higher moments and cumulants of displacement in 
Sec.~\ref{HigherMoments}, and the results are compared 
with simulations. We investigate the parameter dependence 
and time evolution of the probability distribution of 
the net distance from the origin in Sec.~\ref{Probabilities}. 
Moreover, the coupling between longitudinal and 
perpendicular transport is briefly discussed, and 
an analytical expression is obtained for the 
probability that the direction of motion after 
$n$ steps makes an angle $\alpha$ with the current 
direction of motion. Section \ref{Summary} concludes 
the paper.

\section{Model}
\label{Model-2D}
We consider the motion of a self-propelled particle as a 
persistent random walk consisting of steps of different 
lengths and orientations. Here, a two-dimensional motion 
is introduced and the extension to three dimensions is 
discussed in the next section. The stochastic motion is 
described in continuous space and discrete time as follows: 
At each time step, the particle takes a step of length $l$ 
along either its previous direction with probability $p$ 
or a newly chosen direction with probability $s{=}1{-}p$, 
as shown in Fig.~\ref{Fig1}. Thus, $p$ represents the 
self-propulsion of the particle. We assume probability 
distributions $R(\phi)$ and $\mathcal{F}(\ell)$ for 
the rotation angle $\phi{=}\theta{-}\gamma$ and the 
step length $l$ of the walker, respectively. The following 
master equation expresses the evolution of the probability 
density $P_{n}(x,y|\theta)$ for the particle to arrive 
at position $(x,y)$ along the direction $\theta$ at 
time step $n$:
\begin{equation}
\begin{aligned}
&P_{n+1}(x,y|\theta) = p \! \int \!\! d\ell \, \mathcal{F}
(\ell) \, P_{n}\big(x{-}\ell \text{cos}(\theta),y{-}\ell 
\text{sin}(\theta) \big|\theta\big)\\ 
&+ s \!\! \int \!\! d\ell \, \mathcal{F}(\ell) \! 
\int_{-\pi}^{\pi} \!\!\!\! d\gamma \,R(\theta{-}\gamma) 
\, P_{n}\big(x{-}\ell \text{cos}(\theta),y{-}\ell 
\text{sin}(\theta)\big|\gamma\big).
\end{aligned}
\label{Eq:MasterEquation}
\end{equation}                                                                                                                                                                                                                                                                                                 
The terms on the right hand side of the above equation 
correspond to persistent motion with probability $p$ 
and turning with probability $s$. One can obtain 
arbitrary moments of displacement since they are 
accessible by the derivatives of the Fourier transform 
of $P_{n}(x,y|\theta)$, which is defined as
\begin{eqnarray}
P_{n}(\bm\omega|m) \equiv \! \int_{-\pi}^{\pi} \!\!\!\!\! 
d\theta \, e^{im\theta} \! \int \!\! dy \int \!\! dx \,\, 
e^{i\bm\omega\cdot\bm r} P_{n}(x,y|\theta).
\label{Eq:FourierDefinition}
\end{eqnarray}
The arbitrary moment $\langle x^{k_{1}} y^{k_{2}} \rangle$ 
is given by 
\begin{eqnarray}
\begin{aligned}
\langle &x^{k_{1}} y^{k_{2}} \rangle_n \equiv \int \!\! d\theta \int 
\!\! dy \int \!\! dx \,\, x^{k_{1}} y^{k_{2}}  P_{n}(x,y|\theta) \\
&= \left. (-i)^{k_{1}+k_{2}} \frac{\partial^{k_{1}+k_{2}} 
P_{n}(\omega_{x},\omega_{y} |m {=} 0)}{\partial \omega_{x}^{k_{1}}
\partial \omega_{y}^{k_{2}}} \right|_{(\omega_{x},\omega_{y})=(0,0)}.
\end{aligned}
\label{Eq:Moments}
\end{eqnarray}
To study the diffusive behavior of particles one deals 
with the first and second moments of $P_{n}(x,y|\theta)$, 
namely $\langle x\rangle$, $\langle y\rangle$, $\langle 
x^2\rangle$ and $\langle y^2\rangle$. Thus, we first 
focus on the derivation of these quantities in the 
following. The same procedure is followed in 
Sec.~\ref{HigherMoments} to obtain higher moments of 
displacement. A similar Fourier-Z-transform technique 
was applied to study diffusive transport of light in 
foams \cite{FoamRefs}. The first two moments along the 
$x$ direction are given by
\begin{eqnarray}
\begin{aligned}
\langle x \rangle\!_{_n} &\equiv \left. -i \frac{\partial
P_{n}(\omega_{x},\omega_{y} |m {=} 0)}{\partial \omega_{x}
} \right|_{(\omega_{x},\omega_{y})=(0,0)}, \\
\langle x^2 \rangle\!_{_n} &\equiv \left. (-i)^2 \frac{\partial^2
P_{n}(\omega_{x},\omega_{y} |m {=} 0)}{\partial \omega_{x}^2
} \right|_{(\omega_{x},\omega_{y})=(0,0)}.
\end{aligned}
\label{Eq:Moments12}
\end{eqnarray}
Similar expressions can be written for $\langle y\rangle$ 
and $\langle y^2\rangle$. Fourier transforming 
Eq.~(\ref{Eq:MasterEquation}), we find
\begin{eqnarray}
\begin{aligned}
&P_{n+1}(\omega, \alpha |m) = \sum_{k{=}-\infty}^{\infty} 
\Bigg[ i^k e^{-ik\alpha} \times \\
&P_{n}(\omega, \alpha |m{+}k) \big(p+s \,\mathcal{R}(m{+}k)
\big) \! \int \!\! d\ell \, \mathcal{F}(\ell) J_k(\omega\ell) 
\Bigg],
\end{aligned}
\label{Eq:FourierMasterEq11}
\end{eqnarray}
where $J_k(z) {=} \frac{1}{2\pi i^{k}} \int_{-\pi}^{\pi} 
\! d\theta \,\, e^{iz\cos\theta} e^{-ik\theta}$ is the $k$th 
order Bessel's function and $\mathcal{R}(m) {=} \int_{-{\pi}}^{{\pi}} 
\! d\phi \,\, e^{i m \phi} R(\phi)$ is the 
Fourier transform of the rotation angle distribution 
$R(\phi)$. One can expand $P_{n}(\omega, \alpha |m)$ 
as a Taylor series
\begin{eqnarray}
\begin{aligned}
P_{n}(\omega, \alpha |m) 
&= Q_{0,n}(\alpha|m) + i \omega \, \langle \ell \rangle \, 
Q_{1,n}(\alpha|m) \\
& - \frac{1}{2} \omega^2 \, \langle \ell^2 \rangle \, 
Q_{2,n}(\alpha|m)+ \cdot \cdot \cdot,
\end{aligned}
\label{Eq:TaylorExpPw}
\end{eqnarray}
with $\langle \ell \rangle$ and $ \langle \ell^2 \rangle$ 
being the first and second moments of the step-length 
distribution $\mathcal{F}(\ell)$. From Eqs.~(\ref{Eq:Moments12}) 
and (\ref{Eq:TaylorExpPw}) we obtain the moments in terms 
of the Taylor expansion coefficients $Q_{i,n}(\alpha|m)$:
\begin{figure}[t]
\centering
\includegraphics[scale=1.05,angle=0]{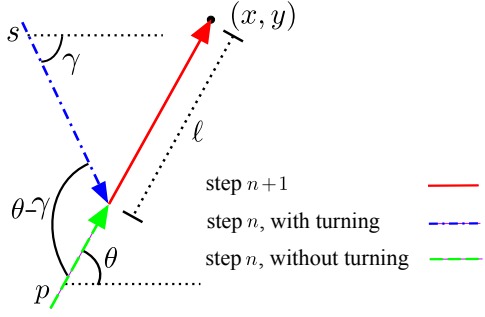}
\caption{(color online). Trajectory of the walker during 
two successive steps.}
\label{Fig1}
\end{figure}
\begin{eqnarray}
\begin{aligned}
\langle x \rangle\!_{_n} &= \! \int \!\! d\ell \, \mathcal{F}(\ell) \, \ell 
\,\, Q_{1,n}(0|0) \,\,\,\,\, = \langle \ell \rangle \, Q_{1,n}(0|0),\\
\langle y \rangle\!_{_n} &= \! \int \!\! d\ell \, \mathcal{F}(\ell) \, \ell 
\,\, Q_{1,n}\big(\frac{\pi}{2}\big|0\big) \,\, = \langle \ell 
\rangle \, Q_{1,n}\big(\frac{\pi}{2}\big|0\big),\\
\langle x^2 \rangle\!_{_n} &= \! \int \!\! d\ell \, \mathcal{F}(\ell) \, 
\ell^2 \,\, Q_{2,n}(0|0) \,\,\, = \langle \ell^2 \rangle \, Q_{2,n}(0|0),\\
\langle y^2 \rangle\!_{_n} &= \! \int \!\! d\ell \, \mathcal{F}(\ell) \, 
\ell^2 \,\, Q_{2,n}\big(\frac{\pi}{2}\big|0\big) = \langle 
\ell^2 \rangle \, Q_{2,n}\big(\frac{\pi}{2}\big|0\big).
\end{aligned}
\label{Eq:MeanMoments}
\end{eqnarray}
By Taylor expansion of both sides of Eq.(\ref{Eq:FourierMasterEq11}), 
one can collect all terms with the same power in $\omega$, leading 
to the following recursion relations for the Taylor coefficients 
$Q_{0,n}(\alpha|m)$, $Q_{1,n}(\alpha|m)$, and $Q_{2,n}(\alpha|m)$:
\begin{eqnarray}
\hspace{-25mm} Q_{0,n+1}(\alpha|m) = Q_{0,n}(\alpha|m) \big(p+s 
\, \mathcal{R}(m)\big), 
\label{Eq:omega0terms3}
\end{eqnarray}

\begin{eqnarray}
\begin{aligned}
\hspace{-8mm} Q_{1,n+1}(\alpha|m) &= Q_{1,n}(\alpha|m) \big(p+s 
\, \mathcal{R}(m)\big)\\
&+ \frac{1}{2} \bigg[ e^{i\alpha} Q_{0,n}(\alpha|m{-}1) \big(p+s 
\, \mathcal{R}(m{-}1)\big)\\ 
&+ e^{-i\alpha} Q_{0,n}(\alpha|m{+}1) 
\big(p+s \, \mathcal{R}(m{+}1)\big) \bigg],
\end{aligned}
\label{Eq:omega1terms4}
\end{eqnarray}
\begin{eqnarray}
\begin{aligned} 
Q_{2,n+1}&(\alpha|m) {=} 
\bigg[ \frac{1}{2} Q_{0,n}(\alpha|m) + Q_{2,n}(\alpha|m) \bigg] 
\big(p+s \, \mathcal{R}(m)\big) \\
&+ \frac{\langle \ell \rangle^2}{\langle \ell^2 \rangle} \bigg[ 
e^{i\alpha} Q_{1,n}(\alpha|m{-}1) \big(p+s \, 
\mathcal{R}(m{-}1)\big)\\
&+ e^{-i\alpha} Q_{1,n}(\alpha|m{+}1) 
\big(p+s \, \mathcal{R}(m{+}1)\big) \bigg]\\
&+ \frac{1}{4} e^{2i\alpha} Q_{0,n}(\alpha|m{-}2) \big(p+s \, 
\mathcal{R}(m{-}2)\big)\\ 
&+ \frac{1}{4} e^{-2i\alpha} Q_{0,n}(\alpha|m{+}2) \big(p+s \, 
\mathcal{R}(m{+}2)\big).
\end{aligned}
\label{Eq:omega2terms4}
\end{eqnarray}

The coupled linear equations (\ref{Eq:omega0terms3}), 
(\ref{Eq:omega1terms4}) and (\ref{Eq:omega2terms4}) 
can be solved by means of the $z$-transform technique. 
The $z$-transform ${G}(z)$ of a function $G_n$ of a 
discrete variable $n=0, 1, 2, \cdot\cdot\cdot$ is 
defined as
\begin{equation}
{G}(z)=\sum_{n=0}^{\infty} G_n z^{-n}.
\label{Eq:zTransformDef1}
\end{equation}
By applying the $z$-transform to 
Eqs.(\ref{Eq:omega0terms3})-(\ref{Eq:omega2terms4}), 
one obtains a set of algebraic equations for 
$Q_{0}(z,\alpha|m)$, $Q_{1}(z,\alpha|m)$ and 
$Q_{2}(z,\alpha|m)$ quantities (see Appendix 
\ref{Appendix1}).

\subsection{The quantities of interest}
\label{QuantitiesOfInterest}
The main goal is to evaluate $\langle x \rangle\!_{_n}$, 
$\langle y \rangle\!_{_n}$, $\langle x^2 \rangle\!_{_n}$, 
and $\langle y^2 \rangle\!_{_n}$, which are given in 
terms of $Q_{i,n}(\alpha|m)$ quantities in 
Eq.~(\ref{Eq:MeanMoments}). Here, we derive $\langle 
x \rangle\!_{_n}$ and $\langle x^2 \rangle\!_{_n}$, 
and a similar approach can be followed to obtain $\langle 
y \rangle\!_{_n}$ and $\langle y^2 \rangle\!_{_n}$. 
The $z$-transform of Eq.~(\ref{Eq:MeanMoments}) 
leads to the following expressions in $z$-space 
\begin{eqnarray}
\begin{aligned}
\langle x \rangle (z) &{=} \sum_{n=0}^{\infty} z^{-n} \langle 
\ell \rangle \, Q_{1,n}(0|0) {=} \langle \ell \rangle \, Q_{1}(z,0|0),\\ 
\langle x^2 \rangle (z)&{=} \sum_{n=0}^{\infty} z^{-n} \langle 
\ell^2 \rangle \, Q_{2,n}(0|0) {=} \langle \ell^2 \rangle \, Q_{2}(z,0|0).
\end{aligned}
\label{Eq:zTransformMeanMoments}
\end{eqnarray}
From Eqs.(\ref{Eq:zTransformMeanMoments}), 
(\ref{Eq:Q1zTransform5}), and (\ref{Eq:Q2zTransform4}), 
one obtains the first and second moments of $x$ in the 
$z$-space (See Eqs.(\ref{Eq:zTransformMeanX-2}) and 
(\ref{Eq:zTransformMeanX2-2}) in Appendix \ref{Appendix2}). 
The last step to get the moments $\langle x \rangle$ 
and $\langle x^2 \rangle$ in real time is the inverse 
$z$-transforming of the $z$-space moments [i.e.\ 
Eqs.(\ref{Eq:zTransformMeanX-2}) and (\ref{Eq:zTransformMeanX2-2})]. 
By introducing $A_i{=}p{+}s\,\mathcal{R}(i)$, 
the resulting moments are:

\begin{eqnarray}
&\hspace{3mm}\langle x \rangle\!_{_n} = \,\,\langle \ell \rangle \, Q_{1,n{=}0}(0|0) 
\hspace{35mm} \\ \nonumber
&{+} \frac{\langle \ell \rangle}{2} Q_{0,n{=}0}(0|-1) A_{_-1}
\frac{1-A_{_-1}^n}{1-A_{_-1}} \hspace{1mm} \\ \nonumber
&{+} \frac{\langle \ell \rangle}{2} Q_{0,n{=}0}(0|1) A_{_1}
\frac{1-A_{_1}^n}{1-A_{_1}}, \hspace{7.5mm} 
\label{Eq:InverseZTransform-X-2}
\end{eqnarray}

\begin{widetext}
\begin{eqnarray}
\begin{aligned}
\hspace{-15mm}\langle x^2 \rangle\!_{_n} &= \,\,\,\langle \ell^2 \rangle Q_{2,n{=}0}(0|0) 
+ n \frac{\langle \ell^2 \rangle}{2} Q_{0,n{=}0}(0|0)\\
&\,\,\,\,\,+ \!\langle \ell \rangle^2 Q_{1,n{=}0}(0|-1) A_{_-1}
\frac{1-A_{_-1}^n}{1-A_{_-1}} +\langle \ell \rangle^2 Q_{1,n{=}0}(0|1) A_{_1}
\frac{1-A_{_1}^n}{1-A_{_1}}\\
&\,\,\,\,\,+ \!\frac{\langle \ell \rangle^2}{2} Q_{0,n{=}0}(0|-2) \,A_{_-1} \, A_{_-2} \times 
\frac{A_{_-1}^{n}A_{_-2} +A_{_-2}^{n}\big(A_{_-1}{-}1\big) + A_{_-2} - 
A_{_-1}}{\big(A_{_-1}{-}A_{_-2}\big) 
\, \big(A_{_-2}{-}1\big) \, \big(1{-}A_{_-1}\big)}\\
&\,\,\,\,\,+ \!\frac{\langle \ell \rangle^2}{2} Q_{0,n{=}0}(0|2)\,A_{_1} \, A_{_2} \times 
\frac{A_{_1}^{n}A_{_2} +A_{_2}^{n}\big(A_{_1}{-}1\big) + A_{_2} - 
A_{_1}}{\big(A_{_1}{-}A_{_2}\big) 
\, \big(A_{_2}{-}1\big) \, \big(1{-}A_{_1}\big)} \\
&\,\,\,\,\,+ \!\frac{\langle \ell \rangle^2}{2} Q_{0,n{=}0}(0|0) \bigg[\, A_{_-1} \, 
\frac{A_{_-1}^{n} - A_{_-1}n + n-1}{(1{-}A_{_-1})^2}
+ \, A_{_1} \, 
\frac{A_{_1}^{n} - A_{_1}n + n-1}{(1{-}A_{_1})^2}\bigg]\\ 
&\,\,\,\,\,+ \!\frac{\langle \ell^2 \rangle}{4} Q_{0,n{=}0}(0|-2)  A_{_-2}
\frac{1{-} A_{_-2}^n}{1{-} A_{_-2}} 
+ \frac{\langle \ell^2 \rangle}{4} Q_{0,n{=}0}(0|2)  A_{_2}
\frac{1{-} A_{_2}^n}{1{-} A_{_2}}.
\end{aligned}
\label{Eq:InverseZTransform-X2-2}
\end{eqnarray}
\end{widetext}

\subsection{Isotropic initial condition}
\label{IsotropicInitialCondition}
For the isotropic initial condition $P_{0}(x,y|\theta) {=} 
\frac{1}{2\pi} \delta(x) \delta(y)$, one finds 
from Eq.~(\ref{Eq:FourierDefinition}) that $P_{0}(\bm\omega|m) 
{=} \frac{1}{2\pi} \int_{-\pi}^{\pi} d\theta \, 
e^{im\theta} {=} \frac{\text{sin}(m\pi)}{m\pi}$. Then, using 
the expansion equation (\ref{Eq:TaylorExpPw}) for $\omega{=}0$, 
it can be seen that the only nonzero $Q$ quantity is 
$Q_{0,n{=}0}(\alpha|0){=}1$ (for $m {=} 0$). Therefore, 
Eq.~(\ref{Eq:InverseZTransform-X-2}) leads to 
\begin{equation}
\langle x \rangle\!_{_n} = 0,
\label{Eq:InverseZTransform-X-IsotropicInit}
\end{equation}
and Eq.~(\ref{Eq:InverseZTransform-X2-2}), after replacing $s$ 
with $1{-}p$, reads
\begin{eqnarray}
\begin{aligned}
&\langle x^2 \rangle\!_{_n} = \\
&n \frac{\langle \ell \rangle^2}{2} 
\bigg[\lambda + \frac{\big(p{+}\mathcal{R}(1) 
{-}p\mathcal{R}(1)\big)}{\big(1{-} p\big)\big(1 
{-}\mathcal{R}(1)\big)} + \frac{\big(p{+} 
\mathcal{R}(-1){-}p\mathcal{R}(-1) \big)}{
\big(1{-}p\big)\big(1{-}\mathcal{R}(-1)\big)}\bigg] \\ 
&\!\!+ \!\frac{\langle \ell \rangle^2}{2} \frac{\big(p {+} 
\mathcal{R}(1){-} p\mathcal{R}(1)
\big)}{\big(1{-}p\big)^2\big(1{-}\mathcal{R}(1) 
\big)^2}\bigg[\big(p{+}\mathcal{R}(1) {-} 
p\mathcal{R}(1)\big)^n{-}1\bigg]\\
&\!\!+ \!\frac{\langle \ell \rangle^2}{2} \frac{\big(p {+} 
\mathcal{R}(-1) {-} p\mathcal{R}(-1) 
\big)}{\big(1{-}p\big)^2\big(1{-}\mathcal{R}(-1) 
\big)^2} \bigg[\big(p{+}\mathcal{R}(-1) {-} 
p\mathcal{R}(-1)\big)^n{-}1\bigg],
\end{aligned}
\label{Eq:InverseZTransform-r2-IsotropicInit}
\end{eqnarray}
with $\lambda{=}\langle \ell^2 \rangle / \langle 
\ell \rangle^2$ being the relative variance of the 
step-length distribution. The $y$-component of the 
mean square displacement $\langle y^2 \rangle_n$ 
has the same form as shown in 
Eq.~(\ref{Eq:InverseZTransform-r2-IsotropicInit}) 
due to symmetry. Therefore, one obtains 
$\langle r^2 \rangle_n {=} \langle x^2 \rangle_n 
{+} \langle y^2 \rangle_n {=} 2 \langle x^2 \rangle_n$. 

\subsection{Long-time behavior}
\label{LongTimeBehavior}
From Eq.~(\ref{Eq:InverseZTransform-r2-IsotropicInit}) in 
the limit of long time (i.e.\ $n {\rightarrow} \infty$) 
one obtains the asymptotic mean square displacement as
\begin{equation}
\begin{split}
&\langle r^2 \rangle_n / \langle \ell \rangle^2 \simeq \\
&n \bigg[\lambda + \frac{\big(p{+}\mathcal{R}(1) 
{-}p\mathcal{R}(1)\big)}{\big(1{-} p\big)\big(1 
{-}\mathcal{R}(1)\big)} + \frac{\big(p{+} 
\mathcal{R}(-1){-}p\mathcal{R}(-1) \big)}{
\big(1{-}p\big)\big(1{-}\mathcal{R}(-1)\big)}\bigg] \\ 
&- \frac{\big(p {+} 
\mathcal{R}(1){-} p\mathcal{R}(1)
\big)}{\big(1{-}p\big)^2\big(1{-}\mathcal{R}(1) 
\big)^2} - \frac{\big(p {+} 
\mathcal{R}(-1) {-} p\mathcal{R}(-1) 
\big)}{\big(1{-}p\big)^2\big(1{-}\mathcal{R}(-1) 
\big)^2}.
\end{split}
\label{Eq:r2-longtime}
\end{equation}
Assuming that the particle moves with a constant 
speed $v$ during the ballistic parts of motion, the 
elapsed time after $n$ steps is $\tau = n \langle 
\ell \rangle /v$. The diffusion constant $D$ is 
related to $\langle r^2 \rangle_n$ as
\begin{eqnarray}
\langle r^2 \rangle_n &=& 4 D \tau,
\label{Eq:DiffusionConstantDef}
\end{eqnarray}
thus, we find
\begin{equation}
D {=} \frac{v\langle \ell \rangle}{4} \bigg[\lambda 
+ \frac{\big(p{+}\mathcal{R}(1) 
{-}p\mathcal{R}(1)\big)}{\big(1{-} p\big)\big(1 
{-}\mathcal{R}(1)\big)} + \frac{\big(p{+} 
\mathcal{R}(-1){-}p\mathcal{R}(-1) \big)}{
\big(1{-}p\big)\big(1{-}\mathcal{R}(-1)\big)}\bigg].
\label{Eq:D}
\end{equation}

\subsection{Turning with left-right symmetry}
\label{LRsymmetry}
The distribution $R(\phi)$ reflects to what extent the 
directions of the successive steps are correlated. The 
analytical method presented in this section allows us to 
handle an arbitrary function $R(\phi)$, however, we are 
particularly interested in the distributions with equal 
probabilities to turn clockwise or anticlockwise. This 
implies that $\mathcal{R}(1) {=} \mathcal{R}(-1) ({\equiv} 
\mathcal{R})$. The asymmetry of the turning angle with 
respect to the arrival direction is quantitatively 
reflected in the value of $\mathcal{R}$ which ranges 
between $-1$ and $1$, with zero denoting a uniform 
case and negative (positive) values corresponding 
to a higher chance of motion to the near backward 
(forward) directions in the next step. When left-right 
symmetry holds the imaginary part of $\mathcal{R}(m)$ 
vanishes, thus, $\mathcal{R}$ becomes
\begin{equation}
\mathcal{R} = \int_{-{\pi}}^{{\pi}} \!\!\! d\phi \,\,
\cos(\phi) \,\, R(\phi),
\label{Eq:R}
\end{equation}
and Eq.~(\ref{Eq:InverseZTransform-r2-IsotropicInit}) 
reduces to 
\begin{eqnarray}
\begin{aligned}
\langle x^2 \rangle_n = &\frac12 n \langle \ell \rangle^2 
\bigg[\lambda + \frac{2\big(p{+}\mathcal{R} 
{-} p\mathcal{R}\big)}{\big(1{-} p\big)\big(1 
{-}\mathcal{R}\big)} \bigg] \\ 
&\!\!+ \!\langle \ell \rangle^2 \frac{\big(p {+} 
\mathcal{R}{-} p\mathcal{R}
\big)}{\big(1{-}p\big)^2\big(1{-}\mathcal{R} 
\big)^2}\bigg[\big(p{+}\mathcal{R} {-} 
p\mathcal{R}\big)^n{-}1\bigg].
\end{aligned}
\label{Eq:r2-LRsymmetry}
\end{eqnarray}

\section{Extension to three dimensions}
\label{Model-3D}
\begin{figure}[b]
\centering
\includegraphics[scale=0.65,angle=0]{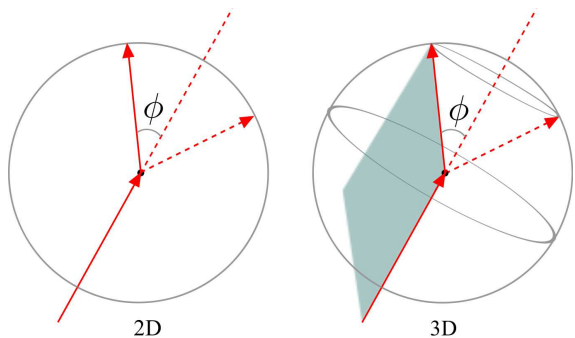}
\caption{(color online). Illustration of symmetric 
rotations with respect to the incoming direction, 
in two (left) and three (right) dimensions.}
\label{Fig2}
\end{figure} 
The analytical approach of Sec.~\ref{Model-2D} can 
be straightforwardly generalized to the persistent 
motion in three dimensions by introducing the 
probability density $P_{n}(x,y,z|\phi,\varphi)$ 
for the particle to arrive at position $(x,y,z)$ 
at time step $n$ along the direction characterized 
by the azimuthal and polar angles $\phi$ and 
$\varphi$, even though the calculations for the 
general motion in three dimensions are quite 
lengthy. However, the processes with symmetric 
turning-angle distribution are of particular 
importance since usually the rotational symmetry 
holds in biological applications. Thus, we 
restrict ourselves in this section to the 
turning-angle distributions with cylindrical 
symmetry with respect to the incoming direction. 
This is the three dimensional analogue to those 
processes in two dimensions which obey left-right 
symmetry. 

\begin{figure}[t]
\centering
\includegraphics[scale=0.37,angle=0]{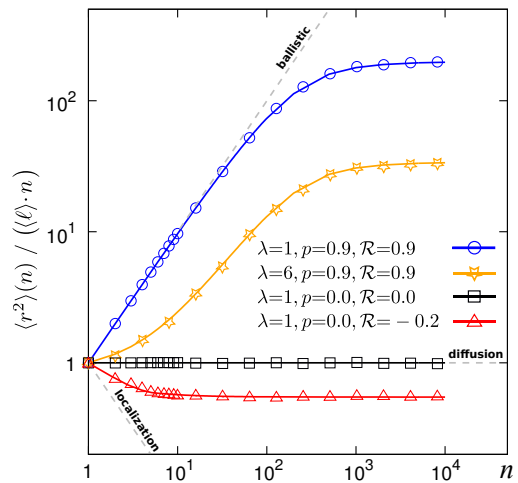}
\caption{(color online). Time evolution of the mean square 
displacement $\langle r^2 \rangle$ scaled by $n$, for 
different values of $\lambda$, $p$, and $\mathcal{R}$. The 
solid lines correspond to analytical predictions via 
Eq.~(\ref{Eq:r2-LRsymmetry}), and the symbols denote 
simulation results.}
\label{Fig3}
\end{figure} 

Similar to the 2D case, we introduce $R(\phi)$ as the 
probability of turning with angle $\phi$ with respect 
to the incoming direction (see Fig.~\ref{Fig2}). The 
polar angle $\varphi$ is supposed to be uniformly 
distributed over the range $[0,2\pi]$. Note that, in 
contrast to the 2D case, the normalization condition 
of $R(\phi)$ in 3D requires an integration over all 
possibilities of $\varphi$ i.e.\ $\int_{0}^{{\pi}} \! R(\phi) \, 
\sin(\phi) \, d\phi {=} 1$. Here again the corresponding 
Fourier transform of $R(\phi)$ is real. Since every 
two successive steps lie in a plane, as shown in 
Fig.~\ref{Fig2}, one can intuitively write the 
same two-dimensional master equation 
[Eq.~(\ref{Eq:MasterEquation})] to describe the 
motion in three dimensions. By solving this master 
equation, one gets a similar expression for $\langle 
x^2 \rangle$ as the 2D solution presented in 
Eq.~(\ref{Eq:InverseZTransform-r2-IsotropicInit}) 
in the case of $\mathcal{R}(1){=}\mathcal{R}(-1)$, 
even though with different prefactors:  
\begin{eqnarray}
\begin{aligned}
\langle x^2 \rangle_n = \, &n \frac{\langle \ell 
\rangle^2}{3} \bigg[\lambda + \frac{2\big(p{+}\mathcal{E} 
{-}p\,\mathcal{E}\big)}{\big(1{-} p\big)\big(1 
{-}\mathcal{E}\big)} \bigg] \\ 
&\!\!+ \!\frac{2\langle \ell \rangle^2}{3} \frac{\big(p {+} 
\mathcal{E}{-} p\,\mathcal{E}
\big)}{\big(1{-}p\big)^2\big(1{-}\mathcal{E} 
\big)^2}\bigg[\big(p{+}\mathcal{E} {-} 
p\,\mathcal{E}\big)^n{-}1\bigg].
\end{aligned}
\label{Eq:x2-3D}
\end{eqnarray}
Here, $\mathcal{E}$ is the real part of the Fourier 
transform of the rotation-angle distribution 
\begin{eqnarray}
\mathcal{E}= \int_{0}^{{\pi}} \!\!\! d\phi \,\,
\cos(\phi) \,\, R(\phi) \, \sin(\phi).
\label{Eq:E}
\end{eqnarray}
Finally, one can obtain the total mean square displacement 
$\langle r^2 \rangle_n = \langle x^2 \rangle_n 
{+} \langle y^2 \rangle_n {+} \langle z^2 
\rangle_n$, which has the same form as in 2D, only 
$\mathcal{R}$ is replaced with $\mathcal{E}$:
\begin{eqnarray}
\begin{aligned}
\langle r^2 \rangle_n = &n \langle \ell \rangle^2 
\bigg[\lambda + \frac{2\big(p{+}\mathcal{E} 
{-} p\mathcal{E}\big)}{\big(1{-} p\big)\big(1 
{-}\mathcal{E}\big)} \bigg] \\ 
&\!\!+ \!\langle \ell \rangle^2 \frac{2\big(p {+} 
\mathcal{E}{-} p\mathcal{E}
\big)}{\big(1{-}p\big)^2\big(1{-}\mathcal{E} 
\big)^2}\bigg[\big(p{+}\mathcal{E} {-} 
p\mathcal{E}\big)^n{-}1\bigg].
\end{aligned}
\label{Eq:r2-LRsymmetry-3D}
\end{eqnarray}

\section{Simulation Results for MSD}
\label{Results}
In this section we compare the analytical predictions 
with the results of extensive Monte Carlo simulations 
obtained from the same step-length $\mathcal{F}(\ell)$ 
and turning-angle $R(\phi)$ distributions, and 
self-propulsion $p$. The formalism introduced in 
Secs.~\ref{Model-2D} and \ref{Model-3D} enables us to 
handle any arbitrary function for $R(\phi)$ and 
$\mathcal{F}(\ell)$, nevertheless, we restrict $R(\phi)$ 
in this section to symmetric distributions along the 
incoming direction for simplicity and because of its 
practical applications in biological systems.

\begin{figure}[t]
\centering
\includegraphics[scale=0.4,angle=0]{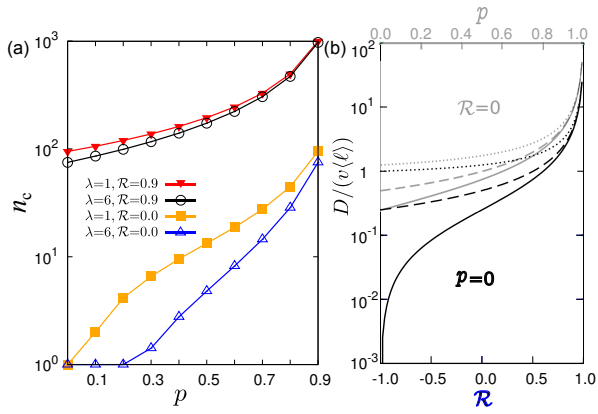}
\caption{(color online). (a) The crossover time $n\!_{_c}$ 
to the asymptotic diffusive regime versus self-propulsion 
$p$. (b) The asymptotic diffusion coefficient $D$ in terms 
of self-propulsion $p$ (light gray curves) or anisotropy $\mathcal{R}$ 
(dark gray curves). The solid, dashed, and dotted lines correspond 
to $\lambda{=}1,2,5$, respectively.}
\label{Fig4}
\end{figure} 
\begin{figure}[b]
\centering
\includegraphics[scale=0.42,angle=0]{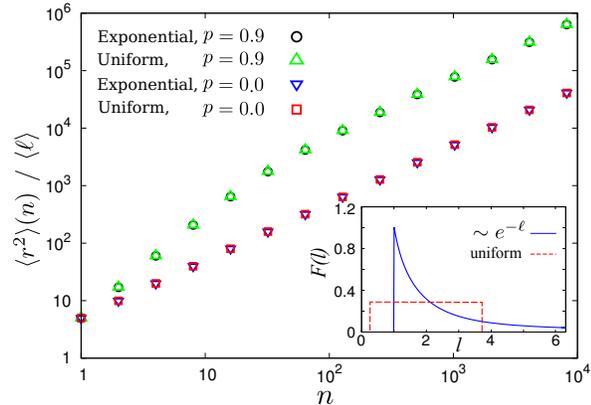}
\caption{(color online). Time evolution of $\langle r^2 
\rangle$ for two different distributions $\mathcal{F}(\ell)$ 
with the same moments $\langle \ell \rangle{=}2$ and 
$\langle \ell^2 \rangle{=}5$ ($\lambda{=}1.25$). Inset: 
Comparison between the exponential distribution $\mathcal{F}(\ell){=} 
e^{1{-}\ell}$ ($\ell\in[1,\infty]$) and the uniform 
distribution $\mathcal{F}(\ell){=}\frac{H(\ell{-}
\ell_\text{min}){+}H(\ell_\text{max}{-}\ell){-}1}
{\ell_\text{max}{-}\ell_\text{min}}$ ($\ell_\text{min}{=}0.268$, 
$\ell_\text{max}{=}3.732$).}
\label{Fig5}
\end{figure} 

We first investigate the overall behavior of the mean square 
displacement for different values of $\lambda$, $p$, and 
$\mathcal{R}$. $\lambda$ is a measure of the heterogeneity 
of the network structure or the diversity of the step sizes, 
and $\mathcal{R}$ quantifies the anisotropy of the structure 
or the asymmetry of the turning angles of the walker. The 
characteristics of $\mathcal{F}(\ell)$ and $R(\phi)$ distributions 
can be considered as the stepping strategy of the random 
walker which may be tunable externally (e.g.\ by controlling 
the external agitation imposed on a driven granular system 
\cite{GranularRefs}) or internally (by controlling the strength, 
density, and spatial arrangement of obstacles in the system, 
or by adjusting the underlying structure of the environment 
such as a porous medium \cite{deAnna13}). $p$ is the self-propulsion 
of the particle (equivalently, the processivity or persistency 
of the walker). The case $p{=}\mathcal{R}{=}0$ and $\lambda{=}1$ 
corresponds to a simple diffusion (see Fig.~\ref{Fig3}). When 
$p$ and $\mathcal{R}$ are both positive, they cooperate to 
send the walker to the near forward directions more frequently, 
resulting in superdiffusion at short time scales. If 
$\mathcal{R}$ is negative, it competes against $p$ which 
may lead to sub, normal, or superdiffusion. At the extreme 
negative value of $\mathcal{R}$ (i.e.\ $\mathcal{R} {\rightarrow} -1$), 
an oscillatory phase can be observed where the particle 
experiences a nearly back and forth motion \cite{Shaebani14}.

\begin{figure}[t]
\centering
\includegraphics[scale=0.38,angle=0]{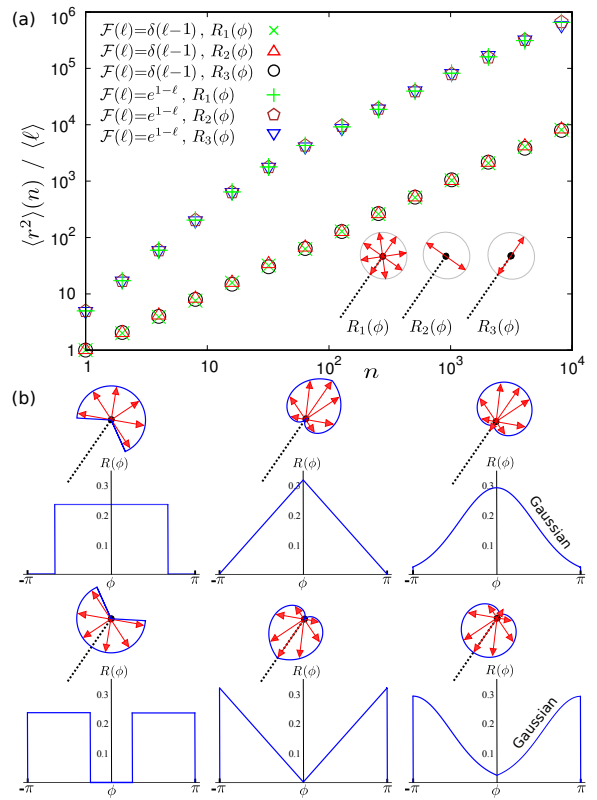}
\caption{(color online). (a) Mean square displacement 
$\langle r^2 \rangle$ vs $n$ for three different 
distributions $R(\phi)$ with $\mathcal{R}{=}0$ given 
in the text. The step-length distribution is chosen 
to be either $\mathcal{F}(\ell){=}\delta(\ell{-}1)$ 
(with $p{=}0$) or $\mathcal{F}(\ell){=}e^{1{-}\ell}$ 
(with $p{=}0.9$). Insets: The possible directions of 
motion at the next step for each $R(\phi)$. (b) A 
few sample distributions $R(\phi)$ with $\mathcal{R}
{\simeq}0.4$ (top) and $\mathcal{R}{\simeq}{-}0.4$ 
(bottom). The dotted lines show the arrival direction 
and the arrows indicate the possible directions in 
the next step, with length being proportional to the 
probability.} 
\label{Fig6}
\end{figure} 

It can be seen from Fig.~\ref{Fig3} that the asymptotic 
behavior of all curves is diffusive. This is due to the 
fact that there is no preferred direction in the system 
and the effective correlations which exist between 
successive step angles are short-range. The crossover 
time $n_c$ to asymptotic diffusion can be estimated by 
balancing the linear and exponential terms in 
Eq.(\ref{Eq:r2-LRsymmetry}). In Fig.~\ref{Fig4}(a), 
$n_c$ is shown as a function of self-propulsion for 
several values of $\mathcal{R}$ and $\lambda$. 
Increasing $p$ and/or $\mathcal{R}$ delays the 
crossover, while the walker gets randomized more 
quickly for strong heterogeneities. The asymptotic 
diffusion coefficient varies by several orders of 
magnitude with control parameters $p$, $\mathcal{R}$, 
and $\lambda$ [see Fig.~\ref{Fig4}(b)].

\begin{figure}[t]
\centering
\includegraphics[scale=0.9,angle=0]{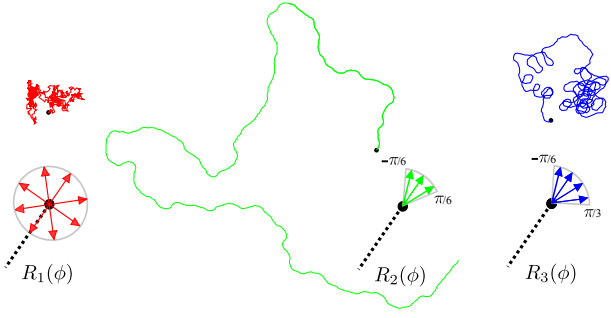}
\caption{(color online). Typical trajectories at $p{=}0$ and 
$\lambda{=}1$ for isotropic (left), forward symmetric (middle), 
and forward asymmetric (right) turning-angle distributions, 
after the same number of steps. The arrows show possible 
directions of motion in the next step.} 
\label{Fig7}
\end{figure} 

A remarkable outcome of the analytical formalism is that the 
anomalous diffusive motion of the particle is fully described 
by the self-propulsion, and the characteristics of the step-length 
and turning-angle distributions, namely the first two moments 
$\langle \ell \rangle$ and $\langle \ell^2 \rangle$ of 
$\mathcal{F}(\ell)$ and the Fourier transform of $R(\phi)$. 
Therefore, one expects that stepping with different distributions 
but with the same key characteristics mentioned above should lead 
to the same results, independent of the functional form of the 
distributions. To verify this finding by simulations, we first 
choose an isotropic distribution $R(\phi){=}\frac{1}{2\pi}$ and 
compare two different step-length distributions with the same 
$\langle \ell \rangle$ and $\langle \ell^2 \rangle$ moments. 
As shown in Fig.~\ref{Fig5}, the simulation results match 
remarkably for an exponential function $\mathcal{F}(\ell){=} 
e^{1{-}\ell}$ and a uniform distribution $\mathcal{F}(\ell)
{=}\frac{H(\ell{-}\ell_\text{min}){+}H(\ell_\text{max}{-}\ell){-}1}
{\ell_\text{max}{-}\ell_\text{min}}$ ($H(x)$ is the Heaviside 
step function), both with $\langle \ell \rangle{=}2$ and $\langle 
\ell^2 \rangle{=}5$.

Next, we choose a given step-length distribution (either 
$\mathcal{F}(\ell){=}\delta(\ell{-}1)$ or $\mathcal{F}
(\ell){=}e^{1{-}\ell}$) and compare three different 
turning-angle distributions: a uniform function $R_1(\phi)
{=}\frac{1}{2\pi}$, a motion restricted to left or right 
directions $R_2(\phi){=}\frac12 (\delta(\phi{-}\pi{/}2){+}
\delta(\phi{-}\pi{/}2))$, and a motion restricted to 
forward or backward directions $R_3(\phi){=}\frac12 
(\delta(\phi){+}\delta(\phi{-}\pi))$. All these 
examples correspond to $\mathcal{R}{=}0$ i.e., on 
average, they have no preference for forward or backward 
motion. Figure~\ref{Fig6}(a) reveals that there is a 
perfect agreement between the simulation results 
obtained for these different turning-angle distributions. 
One can also generate positive or negative values of 
$\mathcal{R}$ from different $R(\phi)$ distributions. 
Several examples are shown in Fig.~\ref{Fig6}(b) for 
$\mathcal{R}{\simeq}0.4$ (or $\mathcal{R}{\simeq}{-}0.4$), 
which all lead to the same diffusive motion.

So far, only symmetric distributions are studied. Now, we 
briefly investigate turning-angle distributions which are 
asymmetric with respect to the incoming direction. Let us 
consider two-dimensional walks for simplicity. An asymmetric 
$R(\phi)$ in this case means that the left-right symmetry 
of turning is broken, leading to (anti-) clockwise spiral 
trajectories. A comparison is made in Fig.~\ref{Fig7} 
between the trajectories obtained from three different 
uniform distributions over the range $(\phi_\text{min},
\phi_\text{max})$: $R_1(\phi)$ is an isotropic function 
corresponding to a normal diffusion ($\mathcal{R}{\equiv}
\mathcal{R}({+}1){=}\mathcal{R}({-}1){=}0$), $R_2(\phi)$ 
is a symmetric function ($\phi_\text{min}{=}{-}\pi{/}6$, 
$\phi_\text{max}{=}\pi{/}6$) which results in $\mathcal{R}
{\equiv}\mathcal{R}({+}1){=}\mathcal{R}({-}1){\simeq}0.95$, 
and $R_3(\phi)$ is an asymmetric distribution over the 
range $[{-}\pi{/}6,\pi{/}3]$ which creates clockwise spirals. 
Here $\mathcal{R}({+}1){\simeq}0.87{+}\text{i}\, 0.23$ 
and $\mathcal{R}({-}1){\simeq}0.87{-}\text{i}\, 0.23$, 
thus, $\mathcal{R}({+}1){\neq}\mathcal{R}({-}1)$. The 
asymptotic diffusion coefficient [Eq.~(\ref{Eq:D})] is 
however a real number $D/v\langle \ell \rangle{=}\frac14 
\big[\lambda{+} 2 \frac{A{-}A^2 {-}B^2}{(1{-}A)^2{+}B^2}\big]$ 
in the absence of self-propulsion, with $A$ and $B$ being 
the real and imaginary parts of $\mathcal{R}({\pm}1)$. 
We obtain $D/ (v\langle \ell \rangle/4) \simeq 1.0$, 
$43.4$, and $2.7$ for $R_1(\phi)$, $R_2(\phi)$, and 
$R_3(\phi)$, respectively, in agreement with the 
simulation results.

\section{Higher moments and cumulants}
\label{HigherMoments}
The procedure described in Sec.~\ref{Model-2D} enables one 
to obtain any arbitrary moment of displacement. To better 
clarify the proposed recipe, we extend the calculations 
to the third and fourth moments in this section, which are 
sufficient to derive up to the fourth cumulants of displacement 
and obtain the skewness and kurtosis of a persistent 
random walk which are measures for the asymmetry and 
peakedness of the probability distribution, respectively. 
We also compare the analytical predictions with Monte Carlo 
simulation results. 

\begin{figure}[b]
\centering
\includegraphics[scale=0.45,angle=0]{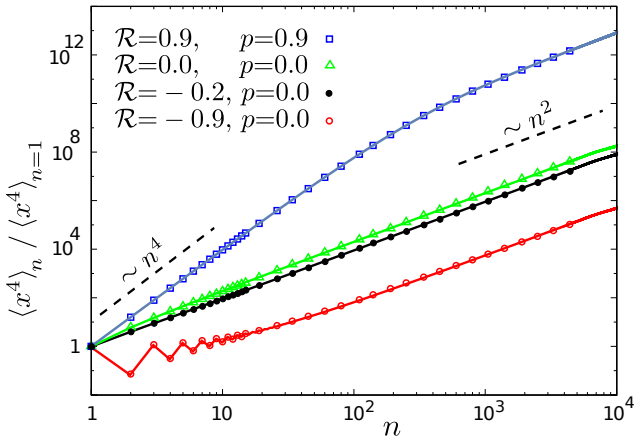}
\caption{(color online). Time evolution of the fourth moment 
of displacement for different values of $\lambda$, $p$, and 
$\mathcal{R}$. The solid lines correspond to analytical 
predictions and the symbols denote simulation results.}
\label{Fig8}
\end{figure}

From Eq.~(\ref{Eq:Moments}), the third and fourth moments 
of the displacement are given by
\begin{eqnarray}
\begin{aligned}
\langle x^3 \rangle\!_{_n} &\equiv \left. (-i)^3 \frac{\partial^3
P_{n}(\omega_{x},\omega_{y} |m {=} 0)}{\partial \omega_{x}^3
} \right|_{(\omega_{x},\omega_{y})=(0,0)},
\end{aligned}
\label{Eq:Moments3}
\end{eqnarray}
and
\begin{eqnarray}
\begin{aligned}
\langle x^4 \rangle\!_{_n} &\equiv \left. (-i)^4 \frac{\partial^4
P_{n}(\omega_{x},\omega_{y} |m {=} 0)}{\partial \omega_{x}^4
} \right|_{(\omega_{x},\omega_{y})=(0,0)}.
\end{aligned}
\label{Eq:Moments4}
\end{eqnarray}
Moreover, by expanding $P_{n}(\omega, \alpha |m)$ up to the forth order 
terms in $\omega$ one finds  
\begin{eqnarray}
\begin{aligned}
P_{n}(\omega, \alpha |m) 
&= Q_{0,n}(\alpha|m) + i \omega \, \langle \ell \rangle \, 
Q_{1,n}(\alpha|m) \\
& - \frac{1}{2} \omega^2 \, \langle \ell^2 \rangle \, 
Q_{2,n}(\alpha|m) \\
& - \frac{i}{6} \omega^3 \, \langle \ell^3 \rangle \, 
Q_{3,n}(\alpha|m) \\
& + \frac{1}{24} \omega^4 \, \langle \ell^4 \rangle \, 
Q_{4,n}(\alpha|m)+ \cdot \cdot \cdot,
\end{aligned}
\label{Eq:TaylorExpPw4thOrder}
\end{eqnarray}
which results in the following relations between $\langle x^3 \rangle$ 
or $\langle x^4 \rangle$ and the Taylor expansion coefficients
\begin{eqnarray}
\begin{aligned}
\langle x^3 \rangle\!_{_n} &= \! \int \!\! d\ell \, \mathcal{F}(\ell) \, 
\ell^3 \,\, Q_{3,n}(0|0) \,\,\, = \langle \ell^3 \rangle \, Q_{3,n}(0|0), \\
\langle x^4 \rangle\!_{_n} &= \! \int \!\! d\ell \, \mathcal{F}(\ell) \, 
\ell^4 \,\, Q_{4,n}(0|0) \,\,\, = \langle \ell^4 \rangle \, Q_{4,n}(0|0).
\end{aligned}
\label{Eq:MeanMomentsX3X4}
\end{eqnarray}
Then, by following the procedure introduced in Sec.~\ref{Model-2D} 
one obtains recursion relations for $Q_{3,n}(\alpha|m)$ and 
$Q_{4,n}(\alpha|m)$ as
\begin{widetext}
\begin{eqnarray}
\nonumber
Q_{3,n+1}(\alpha|m) &{=}& \\ \nonumber
&\,\,& \bigg[ \frac32 \frac{\langle \ell \rangle \langle \ell^2 
\rangle}{\langle \ell^3 \rangle} Q_{1,n}(\alpha|m) {+} 
Q_{3,n}(\alpha|m) \bigg] \big(p{+}s \, \mathcal{R}(m)\big) \\ \nonumber
&+&\bigg[ \frac{\langle \ell \rangle \langle \ell^2 \rangle}
{\langle \ell^3 \rangle} Q_{2,n}(\alpha|m{+}1) + \frac14
Q_{0,n}(\alpha|m{+}1) \bigg] \, \times \frac32 e^{-i\alpha} \big(p{+}s \, \mathcal{R}(m{+}1)\big) \hspace{3.2cm} \\ \nonumber
&+&\bigg[ \frac{\langle \ell \rangle \langle \ell^2 \rangle}
{\langle \ell^3 \rangle} Q_{2,n}(\alpha|m{-}1) + \frac14
Q_{0,n}(\alpha|m{-}1) \bigg] \, \times \frac32 e^{i\alpha} \big(p{+}s \, \mathcal{R}(m{-}1)\big) \\ \nonumber
&+&\bigg[ \frac{\langle \ell \rangle \langle \ell^2 \rangle}
{\langle \ell^3 \rangle} Q_{1,n}(\alpha|m{+}2) \bigg] \frac34 e^{-2i\alpha} 
\big(p{+}s \, \mathcal{R}(m{+}2)\big) \\ \nonumber
&+&\bigg[ \frac{\langle \ell \rangle \langle \ell^2 \rangle}
{\langle \ell^3 \rangle} Q_{1,n}(\alpha|m{-}2) \bigg] \frac34 e^{2i\alpha} 
\big(p{+}s \, \mathcal{R}(m{-}2)\big) \\ \nonumber
&+& \frac18 e^{-3i\alpha} Q_{0,n}(\alpha|m{+}3) \big(p{+}s \, 
\mathcal{R}(m{+}3)\big) \\ 
&+& \frac18 e^{3i\alpha} Q_{0,n}(\alpha|m{-}3) \big(p{+}s \, 
\mathcal{R}(m{-}3)\big),
\label{Eq:omega3terms}
\end{eqnarray}
and
\begin{eqnarray}
\nonumber
Q_{4,n+1}(\alpha|m) &{=}& \\ \nonumber
&\,\,& \bigg[ \frac38 Q_{0,n}(\alpha|m) {+} 3 \frac{\langle 
\ell^2 \rangle^2}{\langle \ell^4 \rangle} Q_{2,n}(\alpha|m) {+} 
Q_{4,n}(\alpha|m) \bigg] \big(p{+}s \, \mathcal{R}(m)\big) \\ \nonumber
&+& e^{-i\alpha} \bigg[ 2 \frac{\langle \ell \rangle \langle \ell^3 \rangle}
{\langle \ell^4 \rangle} Q_{3,n}(\alpha|m{+}1) {+} \frac32 
\frac{\langle \ell \rangle \langle \ell^3 \rangle}{\langle \ell^4 \rangle} 
Q_{1,n}(\alpha|m{+}1)\bigg] \big(p{+}s \, \mathcal{R}(m{+}1)\big) \\ \nonumber
&+& e^{i\alpha} \bigg[ 2 \frac{\langle \ell \rangle \langle \ell^3 \rangle}
{\langle \ell^4 \rangle} Q_{3,n}(\alpha|m{-}1) {+} \frac32 
\frac{\langle \ell \rangle \langle \ell^3 \rangle}{\langle \ell^4 \rangle} 
Q_{1,n}(\alpha|m{-}1)\bigg] \big(p{+}s \, \mathcal{R}(m{-}1)\big) \\ \nonumber
&+& e^{-2i\alpha} \bigg[ \frac32 \frac{\langle \ell^2 \rangle^2}{\langle 
\ell^4 \rangle} Q_{2,n}(\alpha|m{+}2) + \frac14 Q_{0,n}(\alpha|m{+}2) 
\bigg] \big(p{+}s \, \mathcal{R}(m{+}2)\big) \\ \nonumber
&+& e^{2i\alpha} \bigg[ \frac32 \frac{\langle \ell^2 \rangle^2}{\langle 
\ell^4 \rangle} Q_{2,n}(\alpha|m{-}2) + \frac14 Q_{0,n}(\alpha|m{-}2) 
\bigg] \big(p{+}s \, \mathcal{R}(m{-}2)\big) \\ \nonumber
&+& \frac12 e^{-3i\alpha} \frac{\langle \ell \rangle \langle \ell^3 \rangle}
{\langle \ell^4 \rangle} Q_{1,n}(\alpha|m{+}3) \big(p{+}s \, 
\mathcal{R}(m{+}3)\big) 
+ \frac12 e^{3i\alpha} \frac{\langle \ell \rangle \langle \ell^3 \rangle}
{\langle \ell^4 \rangle} Q_{1,n}(\alpha|m{-}3) \big(p{+}s \, 
\mathcal{R}(m{-}3)\big) \\ 
&+& \frac{1}{16} e^{-4i\alpha} Q_{0,n}(\alpha|m{+}4) \big(p{+}s \, 
\mathcal{R}(m{+}4)\big) 
+ \frac{1}{16} e^{4i\alpha} Q_{0,n}(\alpha|m{-}4) \big(p{+}s \, 
\mathcal{R}(m{-}4)\big).
\label{Eq:omega4terms}
\end{eqnarray}
\end{widetext}
The corresponding algebraic equations for $Q_{3}(z,\alpha|m)$ and 
$Q_{4}(z,\alpha|m)$ after the $z$-transform are given in Appendix 
\ref{Appendix1}. From Eq.~(\ref{Eq:MeanMomentsX3X4}), the third 
and fourth moments of $x$ in the $z$-space can be obtained as
\begin{eqnarray}
\begin{aligned}
\langle x^3 \rangle (z) &{=} \sum_{n=0}^{\infty} z^{-n} \langle 
\ell^3 \rangle \, Q_{3,n}(0|0) {=} \langle \ell^3 \rangle \, Q_{3}(z,0|0),\\ 
\langle x^4 \rangle (z)&{=} \sum_{n=0}^{\infty} z^{-n} \langle 
\ell^4 \rangle \, Q_{4,n}(0|0) {=} \langle \ell^4 \rangle \, Q_{4}(z,0|0).
\end{aligned}
\label{Eq:zTransform-Moments3,4}
\end{eqnarray}
By inserting $Q_{3}(z,0|0)$ and $Q_{4}(z,0|0)$ from Eqs.~(\ref{Eq:Q3zTransform4}) 
and (\ref{Eq:Q4zTransform4}), and inverse $z$-transforming 
of the $z$-space moments, we finally obtain $\langle x^3 
\rangle\!_{_n}$ and $\langle x^4 \rangle\!_{_n}$, which are 
very lengthy equations. However, they fortunately reduce to 
simpler forms when we consider the most interesting cases. 
For example, the isotropic initial condition leads to  
\begin{equation}
\langle x^3 \rangle\!_{_n} = 0,
\label{Eq:InverseZTransform-X3-IsotropicInit}
\end{equation}
and if we further limit the motion to a constant 
step size, $\mathcal{F}(\ell){=}\delta(\ell{-}L)$, 
and turning with left-right symmetry, the resulting 
$\langle x^4 \rangle\!_{_n}$ reads   
\begin{widetext}
\begin{eqnarray}
\nonumber
\displaystyle\langle x^4 \rangle\!_{_n} = &\,\,\,\displaystyle\frac{-3}{8} L^4 \Bigg[ 
-\displaystyle\frac{2 (A_{_1}{+}A_{_2})^2 A_{_2}^{n{+}1}}{(1{-}A_{_2})^2 (A_{_1}{-}A_{_2})^2} + 
\displaystyle\frac{4 A_{_1}^{n+1} (n{+}1) \Big(A_{_1} (3 A_{_1}{+}1)-A_{_2}(A_{_1}{+}3)
\Big)}{(1{-}A_{_1})^3 (A_{_1}{-}A_{_2})} \hspace{50mm} \\ 
\nonumber
&+\displaystyle\frac{4 A_{_1}^{n+1} \Big(-8 (2{+}A_{_1}) A_{_1}^2 A_{_2}{+}(A_{_1} 
(7 A_{_1}{+}4){+}1) A_{_1}^2{+}\big(A_{_1} (3 A_{_1}{+}8){+}1\big) A_{_2}^2
\Big)}{(1{-}A_{_1})^4 (A_{_1}{-}A_{_2})^2} \hspace{58mm} \\ 
\nonumber
&+\displaystyle\frac{(n{+}1)(1{+}A_{_1}) \Big(-3 A_{_1}^2+(A_{_1}{-}3) 
(5 A_{_1}{+}3) A_{_2}{+}12 A_{_1}{+}7\Big)}{(1{-}A_{_1})^3 (1{-}A_{_2})}
-\displaystyle\frac{2 (n+1) (n+2) (1{+}A_{_1})^2}{(1{-}A_{_1})^2} \hspace{36.5mm} \\ 
&+ \displaystyle\frac{2 A_{_1} A_{_2} \left(A_{_1}^3{+}18 A_{_1}{+}24\right) 
{-}A_{_2}^2 (A_{_1}{+}1) \Big(A_{_1} \big((A_{_1}{-}5) A_{_1}{+}23\big){+}5\Big) 
{+}A_{_1} (A_{_1}{+}2) \big((A_{_1}{-}6) A_{_1}{-}10\big){+}10 A_{_2}{-}3}{(1{-}A_{_1})^4 (1{-}A_{_2})^2} 
\Bigg], \hspace{8mm}
\label{Eq:4thMoment-general}
\end{eqnarray}
\end{widetext}
with 
\begin{eqnarray}
\nonumber
A_{_1}&{=}p{+}s\,\mathcal{R}(1){=}p{+}s\displaystyle\!\!
\int_{-{\pi}}^{{\pi}} \!\!\!\!\! d\phi \,\cos(\phi) \, R(\phi), \hspace{30mm} \\ 
\nonumber
A_{_2}&{=}p{+}s\,\mathcal{R}(2){=}p{+}s\displaystyle\!\!
\int_{-{\pi}}^{{\pi}} \!\!\!\!\! d\phi \,\cos(2\phi) \, 
R(\phi). \hspace{30mm} 
\end{eqnarray}
The analytical prediction for $\langle x^4 \rangle\!_{_n}$ via 
Eq.~(\ref{Eq:4thMoment-general}) is in agreement with the 
simulation results as shown in Fig.~\ref{Fig8}. It can be seen 
that $\langle x^4 \rangle\!_{_n}$ contains similar information 
as the MSD concerning the anomalous diffusive motion of the 
persistent walker.

\begin{figure}[b]
\centering
\includegraphics[scale=0.7,angle=0]{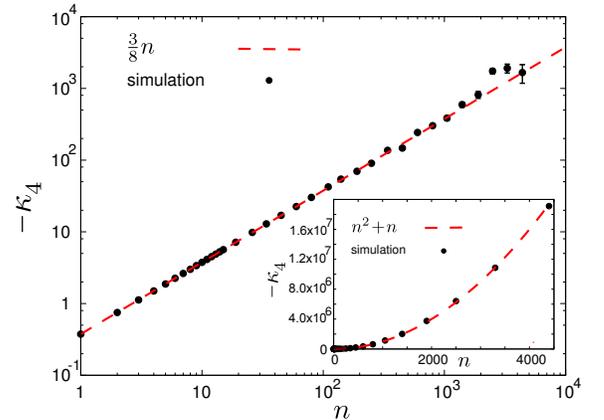}
\caption{The 4th cumulant of $x$ (in units of 
$\langle \ell^4 \rangle$) in terms of $n$ for a random walk 
with $p{=}A_{_1}{=}A_{_2}{=}0$. Inset: The corresponding 4th 
cumulant of the net displacement $r$ vs.\ $n$.} 
\label{Fig9}
\end{figure}
\begin{figure}[t]
\centering
\includegraphics[scale=0.99,angle=0]{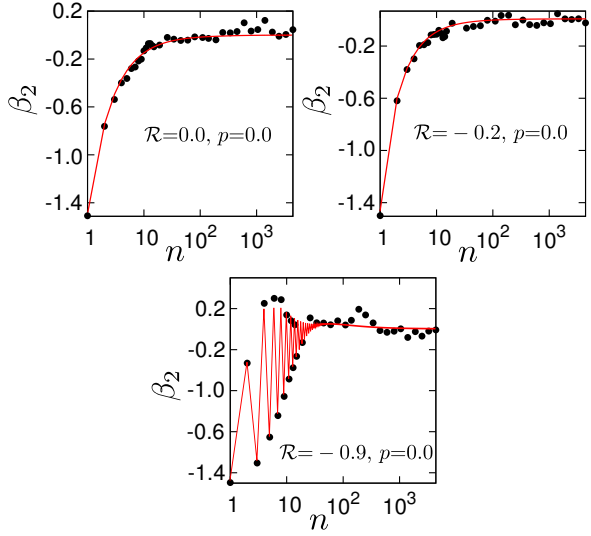}
\caption{The kurtosis measure 
$\beta_2$ in terms of the step number $n$ for different 
turning-angle distributions. The lines correspond to 
analytical predictions via Eq.~(\ref{Eq:SkewnessKurtosis}) 
and the symbols denote simulation results.} 
\label{Fig10}
\end{figure} 

The cumulants are often used in the statistical analysis 
as an alternative to the moments of the distribution. In 
general, the following relations hold between the $n$-th 
cumulant $\kappa_n$ and the moments (shown up to the 4th 
cumulant):
\begin{equation}
\begin{aligned}
\kappa_1 &= \langle x \rangle, \\ 
\kappa_2 &= \langle x^2 \rangle {-} \langle x \rangle^2, \\ 
\kappa_3 &= \langle x^3 \rangle {-} 3 \langle x^2 \rangle 
\langle x \rangle {+} 2 \langle x \rangle^3, \\ 
\kappa_4 &= \langle x^4 \rangle {-} 4 \langle x^3 \rangle 
\langle x \rangle {-} 3 \langle x^2 \rangle^2 {+} 12 \langle 
x^2 \rangle \langle x \rangle^2 {-} 6 \langle x \rangle^4.
\end{aligned}
\label{Eq:CumulantsMoments1}
\end{equation}
If the walker starts from the origin with the isotropic initial 
condition, the odd moments equal zero and the cumulant-moment 
relations reduce to
\begin{equation}
\begin{aligned}
\kappa_1 &= 0, \\ 
\kappa_2 &= \langle x^2 \rangle, \\ 
\kappa_3 &= 0, \\ 
\kappa_4 &= \langle x^4 \rangle {-} 3 \langle x^2 \rangle^2.
\end{aligned}
\label{Eq:CumulantsMoments2}
\end{equation}
In the case of an ordinary random walk with $\mathcal{F}(\ell)
{=}\delta(\ell{-}L)$, we have $p{=}A_{_1}{=}A_{_2}{=}0$, 
$\langle x^2 \rangle / L^2{=}\frac12 n$, and $\langle x^4 
\rangle / L^4 {=}\frac34 n^2 {-} \frac38 n$, which lead to $\kappa_4 / L^4 {=}-\frac38 n$ 
(see Fig.~\ref{Fig9} for comparison with simulation). 
Thus, from Eqs.~(\ref{Eq:r2-LRsymmetry}), (\ref{Eq:4thMoment-general}), 
and (\ref{Eq:CumulantsMoments2}) one can calculate 
the cumulants, from which other useful quantities 
such as the skewness $\beta_1$ and kurtosis $\beta_2$ 
measures can be obtained as
\begin{equation}
\begin{aligned}
\beta_1 &= \frac{\kappa_3}{\kappa_2^{3{/}2}}=0, \\ 
\beta_2 &= \frac{\kappa_4}{\kappa_2^2}=\frac{\langle x^4 \rangle}{\langle 
x^2 \rangle^2} - 3.
\end{aligned}
\label{Eq:SkewnessKurtosis}
\end{equation}
In Fig.~\ref{Fig10}, the time evolution of kurtosis is shown 
for different turning-angle distributions. For a simple random 
walk, $\beta_2$ decreases as $-3/2n$. Moreover, in anomalous 
diffusive cases, $\beta_2$ asymptotically converges to zero 
since the long-term behavior is diffusion.

It is notable that the higher moments are influenced 
by the details of the shape of the turning-angle 
distribution $R(\phi)$. While the MSD depends only on 
$\mathcal{R}{\equiv}\mathcal{R}(1)$ (i.e.\ $\langle \cos 
\phi \rangle$), $\langle x^4 \rangle$ is a function of 
both $\mathcal{R}(1)$ and $\mathcal{R}(2)$ (i.e.\ $\langle 
\cos \phi \rangle$ and $\langle \cos 2\phi \rangle$). 
Thus, looking at the behavior of the higher moments would 
reveal the underlying differences between turning-angle 
distributions, which are not visible from the MSD results. 
For example, $\mathcal{R}$ equals to zero for the three 
different distributions $R(\phi)$ introduced in Fig.~\ref{Fig6}(a), 
thus, their MSD is the same. However, their $\mathcal{R}(2)$ 
is $0$, $-1$, and $1$ for $R_1(\phi)$, $R_2(\phi)$, and 
$R_3(\phi)$, respectively. Therefore, one obtains different 
analytical expressions for their higher moments such as 
$\langle x^4 \rangle$.

\begin{figure*}[t]
\centering
\includegraphics[scale=1.6,angle=0]{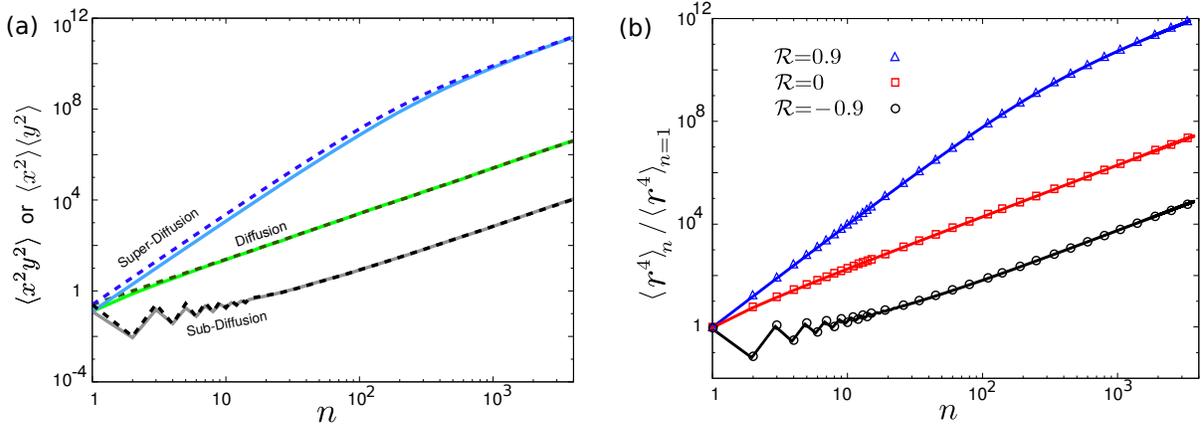}
\caption{(color online). (a) Comparison between 
$\langle x^2 y^2 \rangle$ (solid lines) and $\langle 
x^2 \rangle \langle y^2 \rangle$ (dashed lines) (both 
presented in units of $\langle\ell^4 \rangle$), obtained 
from the simulations where the short-time motion is sub 
($\mathcal{R}{=}-0.9$), normal ($\mathcal{R}{=}0$), or 
superdiffusion ($\mathcal{R}{=}0.9$). (b) $\langle 
r^4 \rangle$ vs.\ $n$ from simulations (symbols) or 
via Eq.~(\ref{Eq:4thMoment-r}) (solid lines).}
\label{Fig11}
\end{figure*}

Besides the components of the displacement, the net 
distance $r$ of the walker from the origin is also 
a quantity of interest. For a persistent walk with an 
arbitrary turning-angle distribution, so far, there has 
been no exact closed-form expression for $\langle r \rangle$. 
For approximate expressions $\langle r \rangle {\approx} 
\sqrt{\langle r^2 \rangle} (1{-}\frac18 \frac{\sigma_{r^2}}
{\langle r^2 \rangle^2})$ (short-time \cite{McCulloch89}) 
and $\langle r \rangle {\approx} \frac12 \sqrt{\pi \langle 
r^2 \rangle}$ (asymptotic \cite{Bovet88}), one deals with 
the calculation of $\langle r^2 \rangle$ and $\langle 
r^4 \rangle$. The second moment $\langle r^2 \rangle$ can 
be obtained from Eq.~(\ref{Eq:r2-LRsymmetry}) since $\langle 
r^2 \rangle{=}\langle x^2 {+} y^2 \rangle{=} \langle x^2 
\rangle {+} \langle y^2 \rangle$. The fourth moment reads 
$\langle r^4 \rangle{=} \langle (x^2 {+} y^2)^2 \rangle 
{=} \langle x^4 \rangle{+}2\langle x^2y^2 \rangle {+} 
\langle y^4 \rangle$. In general, $\langle x^2 y^2 
\rangle {\neq} \langle x^2 \rangle \langle y^2 \rangle$ [see 
Fig.~\ref{Fig11}(a)]. In order to calculate $\langle 
x^2 y^2 \rangle$, one can start from Eq.~(\ref{Eq:Moments}) 
and follow the analytical procedure as explained for 
arbitrary moments $\langle x^i \rangle$. For example, 
a simple random walk with $p{=}A_{_1}{=}A_{_2}{=}0$ 
and $\lambda{=}1$ leads to $\langle x^2 y^2 \rangle 
{=} \frac14 n^2 {-} \frac18 n$ vs.\ $\langle x^2 
\rangle \langle y^2 \rangle {=} \frac14 n^2$. Finally, 
the analytical form of $\langle r^4 \rangle$ for a walker 
with constant step size $L$ and isotropic initial 
conditions is obtained as
\begin{widetext}
\begin{eqnarray}
\nonumber
\displaystyle\langle r^4 \rangle\!_{_n} = \frac14 L^4 \bigg[ \!\! &\!& \frac{(3 
A_{_1}{+}1) n^2{+}(A_{_1}{+}1) (n{-}1) n}{1{-}A_{_1}} 
- 4\frac{A_{_1} A_{_2} \Big(A_{_2} (A_{_1}{+}A_{_2}) 
\left(A_{_1}{-}A_{_1}^{n{-}1}\right){+}(A_{_2}{-}A_{_1}) 
\left(A_{_2}{-}A_{_2}^n\right)\Big)}{(1{-}A_{_1}) 
(1{-}A_{_2}) (A_{_2}{-}A_{_1})^2} \\
\nonumber
&{-}& 4\frac{A_{_1} \left(1{-}A_{_1}^n\right) n {+}A_{_1}^2 
\left(3 n{-}n^2\right){-}2 A_{_1}^2 \left(A_{_1}^n{+}1\right)
{+}A_{_1} (n{-}1)+0.5 A_{_1} (A_{_1}{+}1) (n^2{-}n
{-}2)}{(1{-}A_{_1})^2} \\
\nonumber
&{+}& 4\frac{{-}2 A_{_1}^{n{+}2} n {+} 
A_{_1}^n{+}2 A_{_1}^4{-}A_{_1}^3 \left(n^2{+} 
3 n{-}9\right){-}A_{_1}^2 (n{-}2){+}A_{_1} (n{-}2) 
n}{(1{-}A_{_1})^3} + \frac{2 (A_{_1}{+}1) A_{_2} 
(n{-}1)}{(1{-}A_{_1}) (1{-}A_{_2})}\\
\nonumber
&{+}& 4 \frac{A_{_1}^2 \Big(A_{_1} \big(2 A_{_1} 
(A_{_1}{+}2){+}2 n{-}1\big){-}2 n{+}5\Big){-} 
\Big(A_{_1} \left(2 A_{_1}^3{+}4 A_{_1}{+}3\right)
{+}1\Big) A_{_1}^n}{(1{-}A_{_1})^4} - \frac{2 
(A_{_1}{+}1) \left(A_{_2}^2{-}A_{_2}^{n{+}
1}\right)}{(1{-}A_{_1}) (1{-}A_{_2})^2}\\
\nonumber
&{+}& 8 \frac{\Big(2 A_{_2}{-}A_{_1} (A_{_2}{+}1)\Big) 
\left(A_{_1}^3{-}A_{_1}^{n{+}1}\right) n{-}
A_{_1}^3 (n{-}1) \Big(2 A_{_2}{-}A_{_1} 
(A_{_2}{+}1)\Big)}{(1{-}A_{_1})^3 (A_{_2}{-}A_{_1})} \\
\nonumber
&{-}&\frac{4 A_{_1} A_{_2} (A_{_1}{+}1) 
\Big((1{-}A_{_1}) (n{-}2){-}\left(A_{_1}{-}
A_{_1}^{n{-}1}\right)\Big)}{(1{-}A_{_1})^3 (A_{_2}{-}1)} 
+ 4 \frac{2 \Big(2 A_{_2}{-}A_{_1} (A_{_2}{+}1)\Big) 
A_{_1}^{n{+}1}{+}A_{_1} A_{_2} \left(A_{_1}
{-}A_{_1}^n\right)}{(1{-}A_{_1})^2 (A_{_2}{-}A_{_1})} \\
&{+}&\frac{8 A_{_1} \Big(2 A_{_2}{-}A_{_1} (A_{_2}{+}1)\Big) 
\left(A_{_1}^3{-}A_{_1}^{n{+}2}\right)}{(1{-}A_{_1})^4 
(A_{_2}{-}A_{_1})} + \frac{4 A_{_1} A_{_2}^2 (A_{_1}{+}A_{_2}) 
\left(A_{_2}{-}A_{_2}^{n-1}\right)}{(1{-}A_{_2})^2 (A_{_2}{-}A_{_1})^2} \bigg],
\label{Eq:4thMoment-r}
\end{eqnarray}
\end{widetext}
which is confirmed by the simulation data, as shown 
in Fig.~\ref{Fig11}(b). When $p$ is set to zero, 
Eq.~(\ref{Eq:4thMoment-r}) reduces to the expression 
recently proposed in \cite{Cheung10}, even though our 
formalism allows for obtaining $\langle r^4 \rangle$ 
in the more general case of $p{\neq}0$ and even $\lambda
{\neq}1$ (i.e.\ variable step lengths) and anisotropic 
initial conditions. One can similarly calculate the 
cumulants and relative cumulants such as the kurtosis 
for the net displacement $r$. For example, one finds 
that $-\kappa_4$ grows as $n^2{+}n$ in the simple case 
of $p{=}A_{_1}{=}A_{_2}{=}0$, as shown in Fig.~\ref{Fig9}
(inset). 

\section{Probability distributions of the net distance 
and turning angle after $\bm n$ steps}
\label{Probabilities}
\begin{figure*}
\centering
\includegraphics[scale=1.2,angle=0]{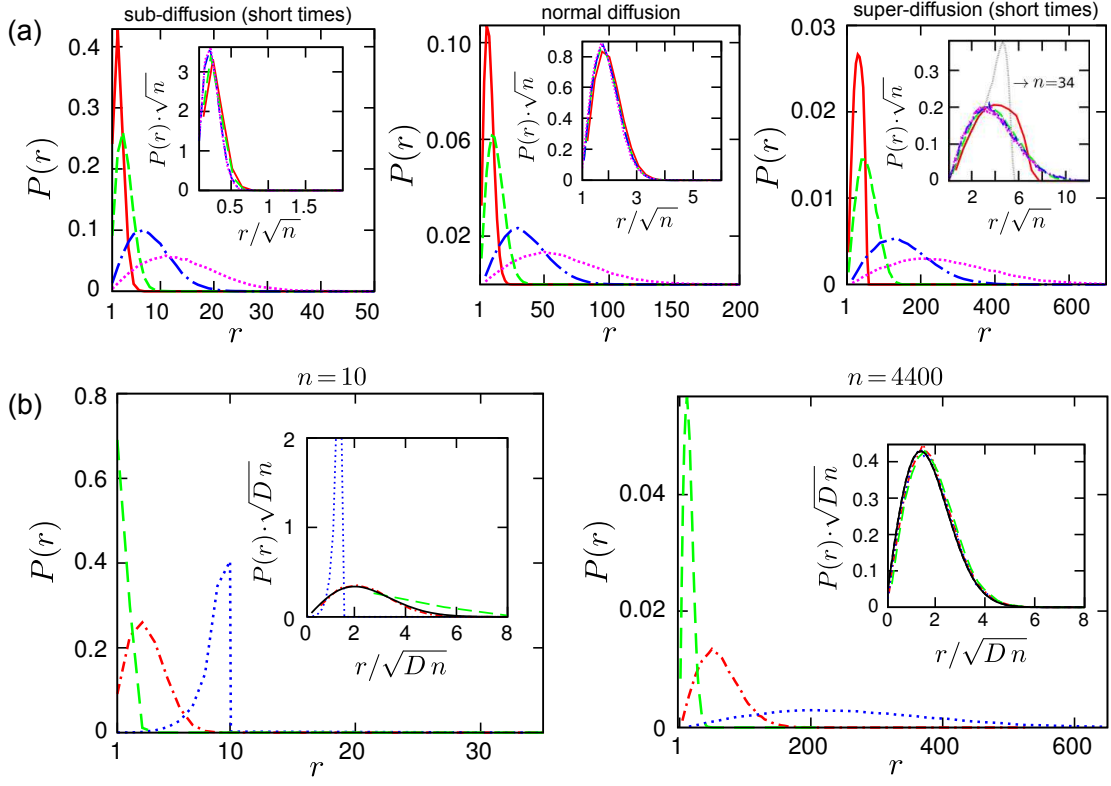}
\caption{(color online). (a) Probability distribution 
of the distance $r$ (in units of $\langle \ell \rangle$) 
from the origin at $n{=}60, 190, 1400,$ and $4400$ 
(solid, dashed, dash-dotted, and dotted lines, respectively), separately 
shown for persistent walks with short-time sub (left), normal 
(middle), and superdiffusive motion (right). Insets: Collapse 
of $P(r)\cdot\sqrt{n}$ vs.\ $r{/}\sqrt{n}$. (b) Comparison 
between persistent walks with short-time sub (dashed lines), 
normal (dash-dotted lines), and superdiffusive motion (dotted lines) 
at the early stages of the walk (left) and after a long time (right). 
Insets: $P(r)\cdot\sqrt{D \, n}$ in terms of $r{/}\sqrt{D \, n}$, 
where $D$ is the asymptotic diffusion coefficient. The solid 
lines are obtained from Eq.~(\ref{Eq:Pr2}).}
\label{Fig12}
\end{figure*}

In this section, we show how the probability densities of the 
position of the random walker and its orientation evolve with 
time. First, we study the probability distribution $P(r)$ of 
the distance $r$ of the persistent random walker from the 
origin in simulations. For a simple random walk in 2D, the 
shape of the distribution at step $n$ approaches 
\begin{eqnarray}
\displaystyle 
P(r) \simeq \frac{2\,r}{\alpha D n} e^{-\frac{r^2}{\alpha D n}}
\label{Eq:Pr}
\end{eqnarray}
in the large 
$n$ limit ($D$ is the diffusion constant and $\alpha{=}4\langle 
\ell \rangle{/}v$). However, the anomalous motion of the persistent 
walker at short times alters the shape of $P(r)$ as well 
as its propagation speed. In figure \ref{Fig12}(a), $P(r)$ 
is plotted at different values of $n$ for $p{=}0$ and three 
turning-angle distributions with $\mathcal{R}=-0.9, 0,$ and 
$0.9$. From Eq.~(\ref{Eq:Pr}) one expects that all normal-diffusion 
data collapse onto a universal curve when $P(r){\cdot}\sqrt{n}$ 
is plotted versus $r{/}\sqrt{n}$ (see the inset). However, the 
distributions of sub and superdiffusion do not follow such 
a master curve at short times. Indeed, $P(r)$ is narrower 
and the peak shifts to the left (right) for subdiffusion 
(superdiffusion) [see Fig.~\ref{Fig12}(b) (left)]. In the extreme 
limit of localization or ballistic motion, $P(r)$ will be a delta 
function at $r{=}0$ or $r{=}n \langle \ell \rangle$, respectively.  
A similar comparison at long times reveals that $P(r)$ broadens 
slower (faster) than a simple random walk in the case of subdiffusion 
(superdiffusion). The shapes are, however, expected to 
follow Eq.~(\ref{Eq:Pr}), as the asymptotic motion is 
diffusive with different diffusion coefficients obtained 
from Eq.~(\ref{Eq:D}). When scaled by $\sqrt{Dn}$, one finds
\begin{eqnarray}
\displaystyle 
P(r){\cdot}\sqrt{Dn} \simeq \frac{2\,r}{\alpha\sqrt{Dn}} 
e^{-\frac{r^2}{\alpha D n}},
\label{Eq:Pr2}
\end{eqnarray}
thus, we achieve a data collapse for $P(r){\cdot}\sqrt{Dn}$ 
vs.\ $r{/}\sqrt{Dn}$ in the asymptotic regime of the persistent 
walks when the motility is purely diffusive, as shown in 
Fig.~\ref{Fig12}(b) (right). For random walks in 3D, 
Eq.~(\ref{Eq:Pr}) should be replaced with $P(r) {\simeq} 
\frac{4}{\sqrt{\pi}} \frac{r^2}{(\alpha D n{/}2)^{3{/}2}} 
e^{-\frac{r^2}{\alpha D n}}$.

\begin{figure*}
\centering
\includegraphics[scale=1.8,angle=0]{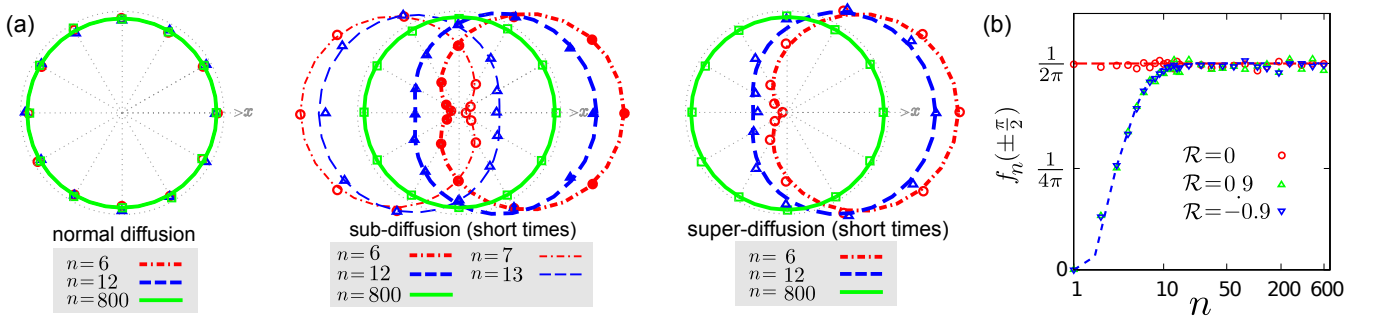}
\caption{(a) Evolution of the angular distribution 
$f\!_{_n}\!(\alpha)$ of the direction of motion $\alpha$ in the 
lab frame. The walker initially arrives along the $+x$ direction. 
A comparison is made between motions with short-time sub 
($\mathcal{R}{=}{-}0.9$), normal ($\mathcal{R}{=}0$), and 
superdiffusion ($\mathcal{R}{=}0.9$). The lines are 
obtained from Eq.~(\ref{Eq:f_alpha2}) and symbols denote 
simulation results. The gray (dotted) lines are guides to eye. 
(b) Probability $f\!_{_n}\!(\pm \frac{\pi}{2})$ of turning 
to a perpendicular direction after $n$ steps, from simulations 
(symbols) or via Eq.~(\ref{Eq:f_alpha2}) (dashed lines).} 
\label{Fig13}
\end{figure*} 

Finally, we investigate how the introduced angular correlations 
between successive steps weaken over time. This is reflected 
e.g.\ in the evolution of the shape of the probability 
$f\!_{_n}\!(\alpha)$ that the direction of motion after 
$n$ steps makes an angle $\alpha$ with the current 
direction of motion. An ordinary random walk is memoryless 
meaning that the walker gets randomized immediately, thus, 
$f\!_{_n}\!(\alpha)$ remains isotropic over the whole range 
of time. However, $f\!_{_n}\!(\alpha)$ is expected to exhibit 
anisotropic shapes at short times for persistent walks, with 
a gradual transition towards isotropic distributions in the 
limit of large $n$. Let us consider the case $p{=}0$ for 
simplicity, i.e.\ the walker turns to a new direction at 
each step. The probability $f\!_{_n}\!(\alpha)$ reads
\begin{equation}
\displaystyle 
f\!_{_n}\!(\alpha) {=} \!\! \int \!\! {\cdot\!\cdot\!\cdot} \!\! 
\int \!\! d\phi_1 {\cdot\!\cdot\!\cdot} d\phi_n \, 
R(\phi_1) {\cdot\!\cdot\!\cdot} R(\phi_n) \delta(\phi_1{+}
{\cdot\!\cdot\!\cdot}{+}\phi_n{-}\alpha),
\label{Eq:f_alpha1}
\end{equation}
where $\phi_i$ denotes the turning angle at step $i$. Using 
the discrete Fourier transform of the delta function 
$\delta(x{-}\alpha){=}\frac{1}{2\pi}\sum_{k{=}-\infty}^{\infty}
e^{ik(x{-}\alpha)}$ one obtains the following expression 
for the walks which are symmetric with respect to the 
arrival direction
\begin{equation}
\displaystyle 
f\!_{_n}\!(\alpha) {=} \frac{1}{2\pi} \Big[ 1{+}2 \!
\sum_{k{=}1}^{\infty} \mathcal{R}^n(k) \, \cos(k\alpha) \Big].
\label{Eq:f_alpha2}
\end{equation}
In the case of $\alpha{=}{\pm}\pi{/}2$, $f\!_{_n}\!({\pm}
\pi{/}2)$ reflects the chance of turning to a perpendicular 
direction after $n$ steps, which can be considered as a 
measure of the coupling between longitudinal and perpendicular 
transport. Figure \ref{Fig13} shows that Eq.~(\ref{Eq:f_alpha2}) 
is in agreement with simulation results.

\section{Summary and outlook}
\label{Summary}
A persistent random walk model was introduced to study 
the stochastic motion of self-propelled particles. By 
developing a general master equation formalism and a 
Fourier-Z-transform technique it was shown that analytical 
exact expressions can be obtained for the time evolution 
of arbitrary moments of displacement. The combination of 
self-propulsion and characteristics of step-size and 
turning-angle distributions lead to a rich transport 
phase diagram at short times. The long-time behavior 
is, however, diffusive since the successive step angles 
in the proposed master equation are only indirectly 
correlated over a few number of steps. This defines 
a time-scale between two arbitrarily chosen steps 
beyond which the steps are practically independent 
of each other. It will be interesting to enhance the 
correlation range e.g.\ by introducing (anti-)cross 
correlations between processivity, step sizes, and 
turning angles. For particular functional forms of 
(anti-)cross correlations, one could even obtain a 
stationary increment for the mean square displacement 
(either sub or superdiffusion) over finite time scales 
as observed for the motion in viscoelastic environments.

\begin{acknowledgments}
This work was funded by the Deutsche Forschungsgemeinschaft 
(DFG) through Collaborative Research Centers SFB 1027 (Projects 
A7 and A3).
\end{acknowledgments}

\appendix

\begin{widetext}
\section{Algebraic coupled equations for the Taylor expansion coefficients $Q_{i}(z,\alpha|m)$}
\label{Appendix1}
\begin{eqnarray}
\hspace{-132mm} Q_{0}(z,\alpha|m) = \frac{z \, 
Q_{0,n{=}0}(\alpha|m)}{z-\big(p 
{+} s \, \mathcal{R}(m)\big)},
\label{Eq:Q0zTransform6}
\end{eqnarray}
\begin{eqnarray}
\begin{aligned}
\hspace{-15mm}Q_{1}(z,\alpha|m) = &\frac{z \, 
Q_{1,n{=}0}(\alpha|m)}{z-\big(p {+} s \, \mathcal{R}(m)\big)} {+} \frac12 e^{i\alpha} 
\frac{Q_{0}(z,\alpha|m{-}1) \big(p{+}s \, \mathcal{R}(m{-}1)\big)}{z-\big(p {+} s \, \mathcal{R}(m)\big)} 
{+} \frac12 e^{-i\alpha} \frac{Q_{0}(z,\alpha|m{+}1) \big(p{+}s \, \mathcal{R}(m{+}1)\big)}{z-\big(p 
{+} s \, \mathcal{R}(m)\big)},
\end{aligned}
\label{Eq:Q1zTransform5}
\end{eqnarray}
\begin{eqnarray}
\begin{aligned}
\hspace{-80mm}Q_{2}(z,\alpha|m) = &\frac{z \, Q_{2,n{=}0}(\alpha|m)}{z-\big(p {+} s \, \mathcal{R}(m)\big)} 
+ \frac12 \frac{Q_{0}(z,\alpha|m) \big(p {+} s \, \mathcal{R}(m)\big)}{z-\big(p {+} s \, \mathcal{R}(m)\big)} \\
&\,\,\,\,\,+ \!\frac{\langle \ell \rangle^2}{\langle \ell^2 \rangle} e^{i\alpha} \frac{Q_{1}(z,\alpha|m{-}1) \big(p {+} s \, \mathcal{R}(m{-}1)\big)}{z-\big(p {+} s \, \mathcal{R}(m)\big)}
+ \!\frac{\langle \ell \rangle^2}{\langle \ell^2 \rangle} e^{-i\alpha} \frac{Q_{1}(z,\alpha|m{+}1) \big(p {+} s \, \mathcal{R}(m{+}1)\big)}{z-\big(p {+} s \, \mathcal{R}(m)\big)} \\
&\,\,\,\,\,+ \!\frac14 e^{2i\alpha} \frac{Q_{0}(z,\alpha|m{-}2) \big(p {+} s \, \mathcal{R}(m{-}2)\big)}{z-\big(p {+} s \, \mathcal{R}(m)\big)}
+ \!\frac14 e^{-2i\alpha} \frac{Q_{0}(z,\alpha|m{+}2) \big(p {+} s \, \mathcal{R}(m{+}2)\big)}{z-\big(p {+} s \, \mathcal{R}(m)\big)},\hspace{16mm}
\end{aligned}
\label{Eq:Q2zTransform4}
\end{eqnarray}
\begin{eqnarray}
\begin{aligned}
\hspace{-85mm}Q_{3}(z,\alpha|m) = &\frac{z \, Q_{3,n{=}0}(\alpha|m)}{z-\big(p {+} s \, \mathcal{R}(m)\big)} 
+ \frac32 \frac{\langle \ell \rangle \langle \ell^2 \rangle}{\langle \ell^3 \rangle}
\frac{Q_{1}(z,\alpha|m) \big(p {+} s \, \mathcal{R}(m)\big)}{z-\big(p {+} s \, \mathcal{R}(m)\big)} \\
&+ \frac38 e^{-i\alpha} \frac{Q_{0}(z,\alpha|m{+}1) \big(p {+} s \, \mathcal{R}(m{+}1)\big)}{z-\big(p {+} s \, \mathcal{R}(m)\big)} 
+ \frac38 e^{i\alpha} \frac{Q_{0}(z,\alpha|m{-}1) \big(p {+} s \, \mathcal{R}(m{-}1)\big)}{z-\big(p {+} s \, \mathcal{R}(m)\big)} \\
&+ \frac18 e^{-3i\alpha} \frac{Q_{0}(z,\alpha|m{+}3) \big(p {+} s \, \mathcal{R}(m{+}3)\big)}{z-\big(p {+} s \, \mathcal{R}(m)\big)} 
+ \frac18 e^{3i\alpha} \frac{Q_{0}(z,\alpha|m{-}3) \big(p {+} s \, \mathcal{R}(m{-}3)\big)}{z-\big(p {+} s \, \mathcal{R}(m)\big)} \\
&+ \frac34 e^{-2i\alpha} \frac{\langle \ell \rangle \langle \ell^2 \rangle}{\langle \ell^3 \rangle}
\frac{Q_{1}(z,\alpha|m{+}2) \big(p {+} s \, \mathcal{R}(m{+}2)\big)}{z-\big(p {+} s \, \mathcal{R}(m)\big)} 
+ \frac34 e^{2i\alpha} \frac{\langle \ell \rangle \langle \ell^2 \rangle}{\langle \ell^3 \rangle}
\frac{Q_{1}(z,\alpha|m{-}2) \big(p {+} s \, \mathcal{R}(m{-}2)\big)}{z-\big(p {+} s \, \mathcal{R}(m)\big)} \\
&+ \frac32 e^{-i\alpha} \frac{\langle \ell \rangle \langle \ell^2 \rangle}{\langle \ell^3 \rangle}
\frac{Q_{2}(z,\alpha|m{+}1) \big(p {+} s \, \mathcal{R}(m{+}1)\big)}{z-\big(p {+} s \, \mathcal{R}(m)\big)} 
+ \frac32 e^{i\alpha} \frac{\langle \ell \rangle \langle \ell^2 \rangle}{\langle \ell^3 \rangle}
\frac{Q_{2}(z,\alpha|m{-}1) \big(p {+} s \, \mathcal{R}(m{-}1)\big)}{z-\big(p {+} s \, \mathcal{R}(m)\big)},  
\end{aligned}
\label{Eq:Q3zTransform4}
\end{eqnarray}
\begin{eqnarray}
\begin{aligned}
\hspace{-85mm}Q_{4}(z,\alpha|m) = &\frac{z \, Q_{4,n{=}0}(\alpha|m)}{z-\big(p {+} s \, \mathcal{R}(m)\big)} 
+ \frac38 \frac{Q_{0}(z,\alpha|m) \big(p {+} s \, \mathcal{R}(m)\big)}{z-\big(p {+} s \, \mathcal{R}(m)\big)}
+ 3 \frac{\langle \ell^2 \rangle \langle \ell^2 \rangle}{\langle \ell^4 \rangle}
\frac{Q_{2}(z,\alpha|m) \big(p {+} s \, \mathcal{R}(m)\big)}{z-\big(p {+} s \, \mathcal{R}(m)\big)} \\
&+ \frac14 e^{-2i\alpha} \frac{Q_{0}(z,\alpha|m{+}2) \big(p {+} s \, \mathcal{R}(m{+}2)\big)}{z-\big(p {+} s \, \mathcal{R}(m)\big)} 
+ \frac14 e^{2i\alpha} \frac{Q_{0}(z,\alpha|m{-}2) \big(p {+} s \, \mathcal{R}(m{-}2)\big)}{z-\big(p {+} s \, \mathcal{R}(m)\big)} \\
&+ \frac{1}{16} e^{-4i\alpha} \frac{Q_{0}(z,\alpha|m{+}4) \big(p {+} s \, \mathcal{R}(m{+}4)\big)}{z-\big(p {+} s \, \mathcal{R}(m)\big)} 
+ \frac{1}{16} e^{4i\alpha} \frac{Q_{0}(z,\alpha|m{-}4) \big(p {+} s \, \mathcal{R}(m{-}4)\big)}{z-\big(p {+} s \, \mathcal{R}(m)\big)} \\
&+ \frac32 e^{-i\alpha} \frac{\langle \ell \rangle \langle \ell^3 \rangle}{\langle \ell^4 \rangle}
\frac{Q_{1}(z,\alpha|m{+}1) \big(p {+} s \, \mathcal{R}(m{+}1)\big)}{z-\big(p {+} s \, \mathcal{R}(m)\big)} 
+ \frac32 e^{i\alpha} \frac{\langle \ell \rangle \langle \ell^3 \rangle}{\langle \ell^4 \rangle}
\frac{Q_{1}(z,\alpha|m{-}1) \big(p {+} s \, \mathcal{R}(m{-}1)\big)}{z-\big(p {+} s \, \mathcal{R}(m)\big)} \\
&+ \frac12 e^{-3i\alpha} \frac{\langle \ell \rangle \langle \ell^3 \rangle}{\langle \ell^4 \rangle}
\frac{Q_{1}(z,\alpha|m{+}3) \big(p {+} s \, \mathcal{R}(m{+}3)\big)}{z-\big(p {+} s \, \mathcal{R}(m)\big)} 
+ \frac12 e^{3i\alpha} \frac{\langle \ell \rangle \langle \ell^3 \rangle}{\langle \ell^4 \rangle}
\frac{Q_{1}(z,\alpha|m{-}3) \big(p {+} s \, \mathcal{R}(m{-}3)\big)}{z-\big(p {+} s \, \mathcal{R}(m)\big)} \\
&+ \frac32 e^{-2i\alpha} \frac{\langle \ell^2 \rangle \langle \ell^2 \rangle}{\langle \ell^4 \rangle}
\frac{Q_{2}(z,\alpha|m{+}2) \big(p {+} s \, \mathcal{R}(m{+}2)\big)}{z-\big(p {+} s \, \mathcal{R}(m)\big)} 
+ \frac32 e^{2i\alpha} \frac{\langle \ell^2 \rangle \langle \ell^2 \rangle}{\langle \ell^4 \rangle}
\frac{Q_{2}(z,\alpha|m{-}2) \big(p {+} s \, \mathcal{R}(m{-}2)\big)}{z-\big(p {+} s \, \mathcal{R}(m)\big)} \\
&+ 2 e^{-i\alpha} \frac{\langle \ell \rangle \langle \ell^3 \rangle}{\langle \ell^4 \rangle}
\frac{Q_{3}(z,\alpha|m{+}1) \big(p {+} s \, \mathcal{R}(m{+}1)\big)}{z-\big(p {+} s \, \mathcal{R}(m)\big)} 
+ 2 e^{i\alpha} \frac{\langle \ell \rangle \langle \ell^3 \rangle}{\langle \ell^4 \rangle}
\frac{Q_{3}(z,\alpha|m{-}1) \big(p {+} s \, \mathcal{R}(m{-}1)\big)}{z-\big(p {+} s \, \mathcal{R}(m)\big)},  
\end{aligned}
\label{Eq:Q4zTransform4}
\end{eqnarray}
\end{widetext}

\begin{widetext}
\section{The first two moments of displacement in the $z$-space}
\label{Appendix2}
\begin{eqnarray}
\begin{aligned}
\hspace{-28mm}\langle x \rangle (z) = \sum_{n=0}^{\infty} z^{-n} \langle x \rangle_n =
\frac{z}{z{-}1} \langle \ell \rangle \, Q_{1,n{=}0}(0|0) 
+\frac{z}{z{-}1} \frac{\langle \ell \rangle}{2}\frac{Q_{0,n{=}0}(0|-1) 
A_{_-1}}{z-A_{_-1}} 
+ \frac{z}{z{-}1} \frac{\langle \ell \rangle}{2} 
\frac{Q_{0,n{=}0}(0|1) A_{_1}}{z-A_{_1}},
\end{aligned}
\label{Eq:zTransformMeanX-2}
\end{eqnarray}
\begin{eqnarray}
\begin{aligned}
\hspace{-40mm}\langle x^2 \rangle (z) = \sum_{n=0}^{\infty} z^{-n} \langle x^2 \rangle_n &= \,\,\,\frac{z}{z{-}1} \langle \ell^2 \rangle Q_{2,n{=}0}(0|0)
+ \frac{z}{(z{-}1)^2} \frac{\langle \ell^2 \rangle}{2} Q_{0,n{=}0}(0|0)\\
&\,\,\,\,\,+ \!\frac{z}{z{-}1} \langle \ell \rangle^2 \bigg[ \frac{Q_{1,n{=}0}(0|1) 
\, A_{_1}}{\big(z-A_{_1}\big)}
+ \frac{Q_{1,n{=}0}(0|-1) \, A_{_-1}} {\big(z-A_{_-1}\big)} \bigg] \\
&\,\,\,\,\,+ \!\frac{z}{(z{-}1)^2} \frac{\langle \ell \rangle^2}{2} \bigg[
\frac{Q_{0,n{=}0}(0|0) \, A_{_1}}{\big(z-A_{_1}\big)}
+ \frac{Q_{0,n{=}0}(0|0) \, A_{_-1}}{
\big(z-A_{_-1}\big)} \bigg]\\
&\,\,\,\,\,+ \!\frac{z}{z{-}1} \frac{\langle \ell \rangle^2}{2} \bigg[ \frac{Q_{0,n{=}0}(0|2) 
\, A_{_1} \, A_{_2}}{
\big(z-A_{_1}\big) 
\, \big(z-A_{_2}\big)} + \frac{Q_{0,n{=}0}(0|-2) \, A_{_-1} \, A_{_-2}}{
\big(z-A_{_-1}\big) \, \big(z-A_{_-2}\big)} \bigg] \\
&\,\,\,\,\,+ \!\frac{z}{z{-}1} \frac{\langle \ell^2 \rangle}{4} \bigg[ \frac{Q_{0,n{=}0}(0|2) 
A_{_2}}{
\big(z-A_{_2}\big)} + \frac{Q_{0,n{=}0}(0|-2) A_{_-2}}{
\big(z-A_{_-2}\big)} \bigg].
\hspace{28mm}
\end{aligned}
\label{Eq:zTransformMeanX2-2}
\end{eqnarray}
\end{widetext}


\begin{thebibliography}{10}
\bibitem{Ross08} J. L. Ross, M. Y. Ali, and D. M. 
Warshaw, Curr. Opin. Cell Biol. \textbf{20}, 41 (2008).
\bibitem{Howse07} J. R. Howse et al., Phys. Rev. Lett. 
\textbf{99}, 048102 (2007).
\bibitem{Peruani07} F. Peruani and L. Morelli, 
Phys. Rev. Lett. \textbf{99}, 010602 (2007).
\bibitem{Shaebani14} M. R. Shaebani, Z. Sadjadi, I. M. Sokolov, 
H. Rieger, and L. Santen, Phys. Rev. E \textbf{90}, 030701(R) (2014).
\bibitem{OldRefs} R. Nossal and G. H. Weiss, J. Theor. Biol. 
\textbf{47}, 103 (1974); P. M. Kareiva and N. Shigesada, 
Oecologia \textbf{56}, 234 (1983).
\bibitem{Codling08} E. A. Codling, M. J. Plank, and S. Benhamou, 
J. R. Soc. Interface \textbf{5}, 813 (2008).
\bibitem{FBM} B. B. Mandelbrot and J. W. Van Ness, 
SIAM Rev. \textbf{10}, 422 (1968); E. Lutz, Phys. 
Rev. E \textbf{64}, 051106 (2001).
\bibitem{CTRW} B. D. Hughes, Random Walks and Random 
Environments (Clarendon, Oxford, UK, 1995), Vol. 1; R. 
Metzler and J. Klafter, Phys. Rep. \textbf{339}, 1 (2000).
\bibitem{Reviews} P. C. Bressloff and J. M. Newby, 
Rev. Mod. Phys. \textbf{85}, 135 (2013); F. 
H\"ofling and T. Franosch, Rep. Prog. Phys. 
\textbf{76}, 046602 (2013).
\bibitem{ProcesivityRefs1} M. Y. Ali et al., Proc. Natl. 
Acad. Sci. \textbf{104}, 4332 (2007); K. Shiroguchi and 
K. Kinosita, Science \textbf{316}, 1208 (2007).
\bibitem{Vershinin07} M. Vershinin et al., Proc. Natl. 
Acad. Sci. \textbf{104}, 87 (2007).
\bibitem{ProcesivityRefs2} Y. Okada, H. Higuchi, and N. 
Hirokawa, Nature \textbf{424}, 574 (2003); T. L. 
Culver-Hanlon et al., Nat. Cell Biol. \textbf{8}, 264 (2006).
\bibitem{McCulloch89} C. E. McCulloch and M. L. Cain, 
Ecology \textbf{70}, 383 (1989).
\bibitem{Bovet88} P. Bovet and S. Benhamou, J. Theor. 
Biol. \textbf{131}, 419 (1988).
\bibitem{FoamRefs} M.-F. Miri and H. Stark, J. Phys. A 
\textbf{38}, 3743 (2005); Z. Sadjadi and M.-F. Miri, 
Phys. Rev. E \textbf{84}, 051305 (2011); Z. Sadjadi et al., 
Phys. Rev. E \textbf{78}, 031121 (2008).
\bibitem{GranularRefs} M. R. Shaebani, J. Sarabadani, 
and D. E. Wolf, Phys. Rev. Lett. \textbf{108}, 198001 
(2012); Phys. Rev. E \textbf{88}, 022202 (2013); A. 
Kudrolli, G. Lumay, D. Volfson, and L. S. Tsimring, 
Phys. Rev. Lett. \textbf{100}, 058001 (2008). 
\bibitem{deAnna13} P. de Anna et al., Phys. Rev. Lett. 
\textbf{110}, 184502 (2013).
\bibitem{Cheung10} A. Cheung, J. Theor. Biol. \textbf{264}, 
641 (2010).
\end{thebibliography}
\end{document}